\documentclass[pdflatex,sn-mathphys-num]{sn-jnl}

\usepackage{amsmath,amssymb,amsfonts}
\usepackage{amsthm}

\usepackage{euscript}   % replaces mathrsfs (prevents font warnings)

\usepackage{graphicx}
\usepackage{booktabs}    % KEEP only here
\usepackage{multirow}
\usepackage{adjustbox}   % KEEP only here (but don't use for this table)
\usepackage{rotating}
\usepackage{pdflscape}

\usepackage{algorithm}
\usepackage{algpseudocode}

\usepackage{listings}

\usepackage[title]{appendix}

\usepackage{xcolor}
\usepackage{textcomp}
\usepackage{manyfoot}

\usepackage{booktabs}
\usepackage{adjustbox}

\usepackage{float}

\usepackage{hyperref}
\hypersetup{breaklinks=true}

\makeatletter
\newcount\sn@nullaux \sn@nullaux=\m@ne
\AddToHook{begindocument/before}{\let\sn@realaux\@auxout \let\@auxout\sn@nullaux}
\AddToHook{begindocument/end}{\let\@auxout\sn@realaux}
\renewcommand\NAT@sort@cites[1]{%
  \let\NAT@cite@list\@empty
  \@for\@citeb:=#1\do{\expandafter\NAT@star@cite\@citeb\@@}%
  \@ifnum{\NAT@sort>\z@}{%
    \expandafter\NAT@sort@cites@\expandafter{\NAT@cite@list}%
  }{}%
}%
\renewcommand\nocite[1]{\@bsphack\@esphack}
\makeatother

\theoremstyle{thmstyleone}%
\theoremstyle{thmstyletwo}%

\theoremstyle{thmstylethree}%

\begin{document}

\title[Article Title]{Cardiovascular Digital Twins from Physics Based to Data Driven Approaches}


\author*[1]{\fnm{Emmanuel} \sur{Lwele}}\email{e.lwele@shu.ac.uk}

\author[2]{\fnm{Francis} \sur{Chikweto}}\email{chikweto.francis.b6@tohoku.ac.jp}

%\author[1,3]{\fnm{Erica} \sur{Dall'Armellina}}\email{E.DallArmellina@leeds.ac.uk}

\affil*[1]{\orgdiv{Materials and Engineering Research Institute},
  \orgname{Sheffield Hallam University}, \city{Sheffield},
  \country{United Kingdom}}

\affil[2]{\orgdiv{Medical Engineering and Cardiology Department},
  \orgname{Institute of Development, Aging and Cancer (IDAC), Tohoku University},
  \city{Sendai}, \country{Japan}}

%\affil[3]{\orgdiv{Leeds Institute of Cardiovascular and Metabolic Medicine},
  %\orgname{University of Leeds}, \city{Leeds}, \country{United Kingdom}}

\abstract{Cardiovascular digital twins aim to create patient-specific
computational models that evolve with clinical data to support diagnosis,
prognosis, and therapy optimisation. Mechanistic models provide physiological
interpretability but remain computationally demanding, whereas data-driven
approaches improve scalability yet risk limited robustness. Emerging
physics-informed, graph-based, and hybrid methods integrate physical
constraints with relational learning across vascular networks. We review
modelling paradigms, data assimilation frameworks, validation challenges,
and translational pathways toward clinically deployable cardiovascular
digital twins.}


\maketitle



\maketitle

\section{Introduction: Why cardiovascular digital twins matter}
\label{Introducation}
\subsection{Motivation and clinical need}

\begin{figure}[H]
\centering
\includegraphics[width=0.7\linewidth]{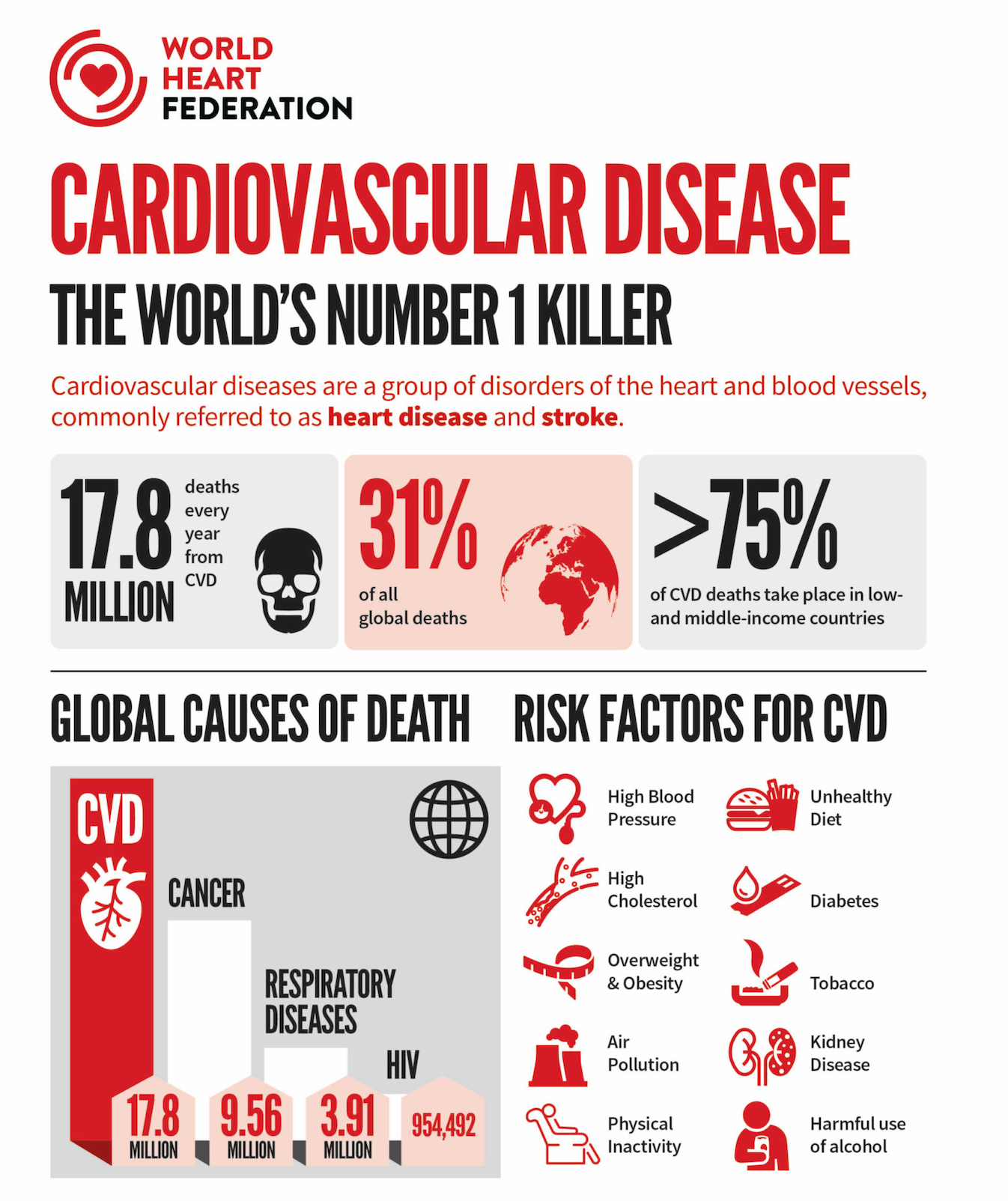}
\caption{Global burden of cardiovascular diseases (CVDs). Source: World Health Organisation (WHO) and partners.}
\label{fig:CVD_WHO}
\end{figure}

Cardiovascular diseases (CVDs) remain the leading cause of global morbidity and mortality, accounting for an estimated 17.8 million deaths annually as of 2025 (Figure~\ref{fig:CVD_WHO}). This persistent burden reflects the intrinsic complexity of cardiovascular physiology and the marked inter-individual variability that characterises disease expression and treatment response. Cardiovascular function emerges from tightly coupled nonlinear interactions spanning multiple spatial and temporal scales, from cellular electrophysiology and myocardial mechanics to vascular hemodynamics, metabolic regulation, and long-term structural remodelling. Disease phenotypes such as heart failure, arrhythmias, and atherosclerosis evolve over prolonged time horizons and are shaped by heterogeneous anatomical, physiological, genetic, and environmental factors. This multiscale complexity poses a fundamental challenge for reliable prediction at the individual level \cite{jeske2020digital,sel2024building}.

Despite advances in cardiovascular imaging, sensing technologies, and interventional therapies, clinical decision-making continues to rely predominantly on population-based risk models and cohort-derived guidelines. These frameworks are calibrated on large datasets and therefore reflect average trajectories rather than patient-specific physiological dynamics. As emphasised in recent reviews of computational cardiology and digital health, such models often fail to account for individual anatomy, boundary conditions, comorbidities, and adaptive responses, limiting their predictive utility for precision cardiology \cite{kissas2023towards,huang2024application}.

The limitations of population-averaged modelling are particularly evident in complex cardiovascular conditions. Patients with apparently similar clinical profiles may experience markedly different responses to identical therapies, underscoring the inadequacy of one-size-fits-all approaches \cite{canino2025artificial,meijer2023digital}. These shortcomings extend to clinical trials, which remain costly and time-consuming and frequently under-represent elderly, multimorbid, or rare-disease populations. Recent work on in silico trials and digital twin–enabled simulation frameworks argues that an exclusive reliance on population-level evidence may obscure mechanistic insight and delay innovation in personalised treatment strategies \cite{coorey2022health,strocchi2025cardiac}.

Collectively, the global burden of cardiovascular disease and the recognised limitations of population-based prediction highlight a translational gap: while large datasets inform guidelines, they do not readily translate into dynamic, patient-specific models capable of real-time forecasting. Addressing this population-to-individual gap requires computational frameworks that integrate mechanistic understanding with continuous patient data and evolve alongside the patient’s physiological state. Cardiovascular digital twins have emerged as a promising response to this need \cite{sel2024building,zhao2025physics}.

\subsection{Digital twins in cardiovascular healthcare}
\label{Digital Twins in Health}

The concept of the digital twin originated in engineering as a virtual replica
of a physical system that remains synchronised with real-world observations.
In healthcare, a digital twin is generally defined as a patient-specific
computational representation that integrates mechanistic models with
longitudinal clinical data to support prediction, monitoring, and
decision-making across the care pathway~\cite{zhang2024concepts,huang2024application}.
Unlike static simulation models, digital twins are dynamic entities designed
to assimilate new data and update their internal states and parameters over
time.

It is useful to distinguish three related concepts along a spectrum of
bidirectionality. A \emph{digital model} is a computational representation of
a physical system that is not connected to real-world data streams. A
\emph{digital shadow} receives data from the physical system to update its
internal state but does not feed information back to influence physical
actions or clinical decisions. A \emph{digital twin}, by contrast, is
characterised by bidirectional coupling: data flows from the patient to the
model for calibration, and model outputs feed back to inform clinical
decisions or interventions. This distinction is not merely terminological; it
delineates the transformative potential of digital twins relative to existing
simulation and decision-support tools~\cite{sel2024building}.

In cardiovascular medicine, digital twins aim to represent the structure and
function of the heart and vasculature in a physiologically grounded and
clinically actionable manner. They typically combine models of hemodynamics,
electrophysiology, and tissue mechanics with patient-specific data derived from
imaging, physiological monitoring, and electronic health
records~\cite{kissas2023towards,sel2024building}. This integration enables the prediction of disease progression, virtual testing of interventions, and
exploration of alternative therapeutic scenarios tailored to the individual.

\begin{figure}[htbp]
\centering
\includegraphics[width=0.9\textwidth]{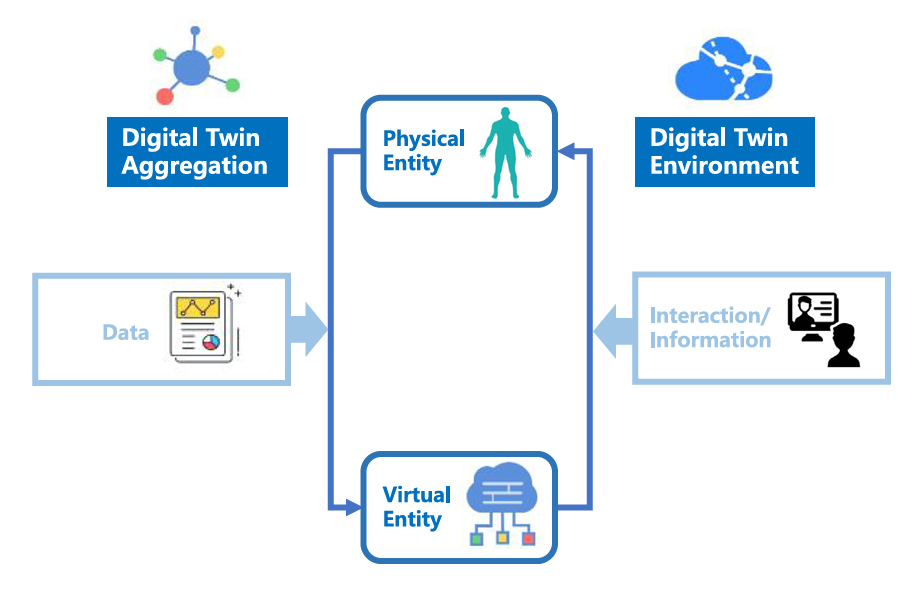}
\caption{Conceptual architecture of a healthcare digital twin, illustrating
bidirectional coupling between the patient and a virtual model. Clinical data
are assimilated to update the twin, while predictions and insights are fed
back to support adaptive, longitudinal, and patient-specific decision making.
Reproduced with permission from Zhang et al.~\cite{zhang2024concepts}.}
\label{fig:dt_architecture}
\end{figure}

Figure~\ref{fig:dt_architecture} illustrates the defining characteristic of a
true digital twin: bidirectional coupling between the patient and the virtual
model. Incoming data are used to recalibrate model states or parameters,
predictions are generated to inform clinical decisions, and observed outcomes
feed back into the twin to refine subsequent predictions. This closed-loop
architecture distinguishes digital twins from conventional computational tools.

Traditional simulation models, including many biophysical cardiovascular
models, are typically constructed offline and executed under fixed assumptions.
Although they provide mechanistic insight, they do not evolve with patient
state and therefore represent static snapshots---analogous to digital
shadows---rather than adaptive representations~\cite{canino2025artificial}.
Decision-support systems form a second category frequently conflated with
digital twins. These systems often rely on statistical or machine learning
models trained on historical datasets to assist with diagnosis or risk
stratification. While valuable for workflow optimisation, they generally lack
explicit physiological modelling and persistent patient-specific
representations, and therefore do not satisfy the defining characteristics of
closed-loop digital twins~\cite{meijer2023digital,coorey2022health}.

The distinction is not merely semantic. The transformative potential of
cardiovascular digital twins lies in their ability to couple mechanistic
understanding with real-time data assimilation, quantify uncertainty, and
support adaptive therapy. Achieving this vision requires advances in model
personalisation, scalable computation, and rigorous validation frameworks,
which collectively define the frontier of digital twin research in
cardiovascular healthcare~\cite{zhang2024concepts,kissas2023towards}.

\subsection{Scope of this review}
\label{Scope of this review}

This review critically evaluates the modelling paradigms that constitute the  foundation of contemporary cardiovascular digital twins and investigates how
diverse computational strategies facilitate personalisation, prediction, and
clinical translation. Adopting a methodology-centric perspective, we perform a
comparative analysis of classical biophysical models, data-driven machine learning approaches, physics-informed neural networks, graph neural networks,
and hybrid modelling frameworks. Rather than providing a disease-specific
inventory of applications, we concentrate on elucidating how modelling
decisions affect interpretability, scalability, data demands, and uncertainty
quantification~\cite{sel2024building,zhang2024concepts}.

Throughout this review, the terms \emph{mechanistic}, \emph{biophysical}, and
\emph{physics-based} are used interchangeably to describe models whose
governing equations are derived from physical and physiological
principles, as opposed to \emph{data-driven} models that learn input--output
mappings from data without explicit physical constraints. The term
\emph{classical} refers to established numerical methods predating the
machine learning era, such as finite element and finite volume solvers.

By synthesising findings across these paradigms, we address four recurrent
questions: (i)~what conceptual and functional features distinguish a
cardiovascular digital twin from conventional computational simulations or
decision-support systems; (ii)~how different modelling strategies negotiate
the trade-off between biophysical fidelity and computational efficiency;
(iii)~how patient-specific data can be integrated and assimilated to enable
continual model updating over time; and (iv)~which validation methodologies
and uncertainty-quantification frameworks are required to support reliable and
ethically acceptable clinical deployment~\cite{canino2025artificial,zhao2025physics}.
Through this integrative analysis, we seek to clarify the current
methodological landscape and to delineate research priorities most likely to
facilitate the development of scalable, interpretable, and clinically
implementable cardiovascular digital twins.

This review adopts a structured narrative approach informed by a targeted
literature search. Searches were conducted across PubMed, Scopus, and Web of
Science, supplemented by Google Scholar for preprints and grey literature.
Search terms included combinations of: \emph{cardiovascular digital twin},
\emph{cardiac computational model}, \emph{physics-informed neural network
heart}, \emph{graph neural network haemodynamics}, \emph{surrogate model
cardiovascular}, \emph{data assimilation cardiac}, \emph{uncertainty
quantification cardiovascular}, and \emph{electromechanical heart model}. The
search was not restricted by date but prioritised publications from 2015
onwards, with seminal earlier works included where foundational. Studies were
selected to achieve balanced coverage across mechanistic, data-driven, and
hybrid paradigms, with emphasis on personalisation, validation, and
translational relevance. Conference proceedings, preprints, and theses were
included selectively where they represent significant methodological
contributions not yet available in peer-reviewed form. No formal PRISMA
screening protocol was applied, consistent with the structured narrative review
design.

\section{Cardiovascular Digital Twins: Definitions and Architecture}
\label{sec:architecture}

\subsection{Digital Twin Framework for Cardiovascular Systems}

A cardiovascular digital twin is defined as a closed-loop framework
comprising three coupled components: (i)~the physical system, representing the
patient's cardiovascular anatomy and physiology; (ii)~a virtual computational
model; and (iii)~a data assimilation and feedback mechanism that synchronises
the model with real-world observations and generates virtual realisations that
translate computational predictions into actionable clinical decisions or
physical interventions.

\begin{figure}[H]
\centering
\includegraphics[width=\linewidth]{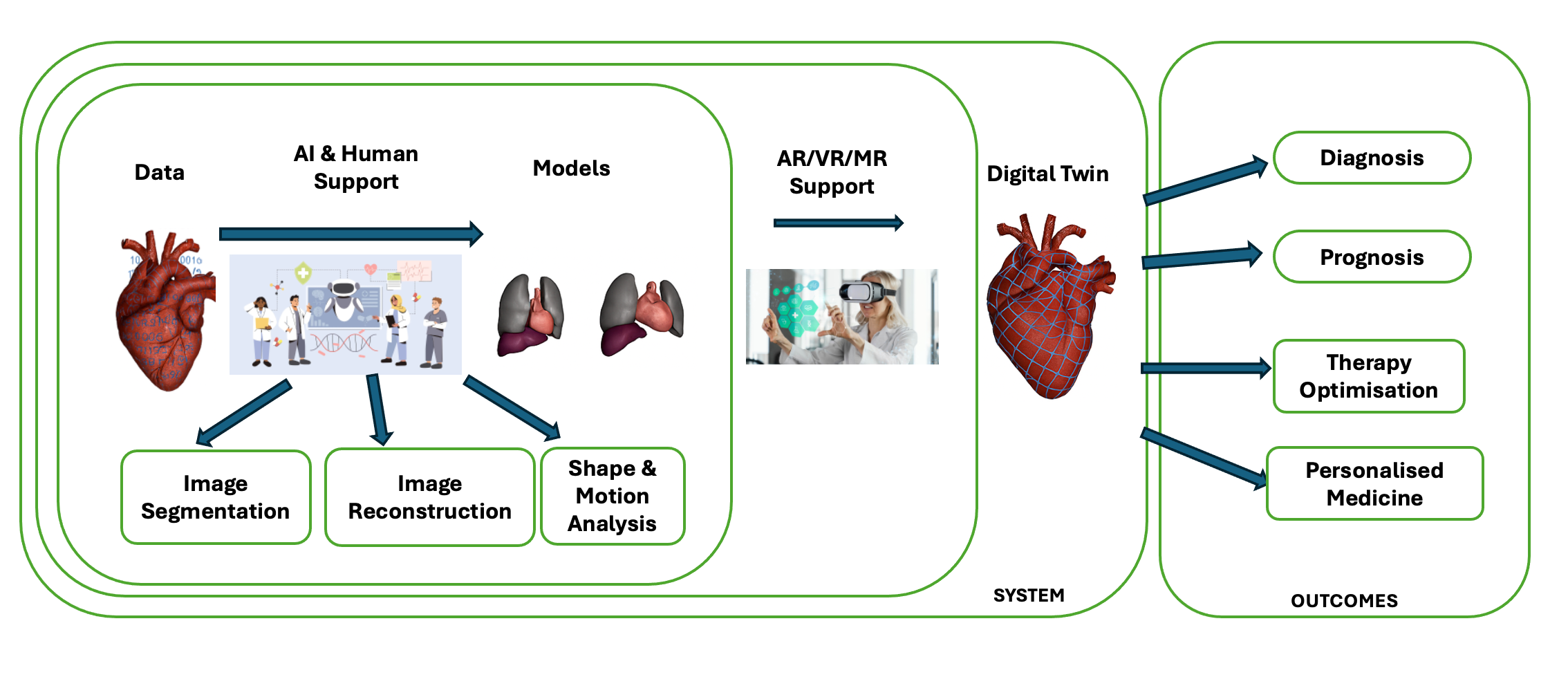}
\caption{Conceptual framework of a cardiovascular digital twin illustrating
the coupling of patient data, AI-assisted modelling, and virtual
representations to support diagnosis, prognosis, therapy optimisation, and
personalised medicine. Note that imaging data are not a universal
prerequisite: certain applications, including 0D haemodynamic monitoring and
continuous non-invasive blood pressure tracking, operate using physiological
signal streams alone, without requiring imaging inputs.}
\label{fig:cv_dt_framework}
\end{figure}

This third component is what distinguishes a digital twin from a digital
shadow: in a shadow, data flow is unidirectional (patient to model), whereas
in a twin, model outputs actively inform and update physical clinical
actions~\cite{sel2024building,zhang2024concepts}. This tripartite architecture
has emerged as a unifying definition across cardiovascular and healthcare
digital twin literature. Figure~\ref{fig:cv_dt_framework} illustrates this
conceptual structure.

The physical system represents a time-varying cardiovascular state,
encompassing cardiac structure, vascular geometry, haemodynamics,
electrophysiology, and disease-related remodelling. These properties evolve in
response to therapy, progression, and lifestyle factors, reinforcing the need
to treat the patient as a dynamic system rather than a static
snapshot~\cite{sel2024building,zhang2024concepts}. Continuous acquisition of
multimodal data enables this temporal representation.

The virtual model acts as a computational surrogate and may be instantiated
through biophysical simulations, data-driven surrogates, or hybrid
formulations. Unlike traditional simulations executed under fixed assumptions,
digital twin models are designed to update states and parameters as new
information becomes available~\cite{canino2025artificial}. AI-assisted
pipelines support image segmentation, geometry reconstruction, and motion
analysis, forming the basis for patient-specific virtual representations.

The defining feature of a digital twin is bidirectional coupling. Incoming
imaging, physiological, or wearable data recalibrate the model through
parameter estimation or learning-based adaptation. Model predictions can then
simulate disease trajectories or treatment scenarios, with outcomes feeding
back into subsequent updates. This closed-loop architecture underpins adaptive
decision support and is widely recognised as essential for predictive validity
and clinical translation~\cite{sel2024building,zhao2025physics}.

\subsection{Levels of Cardiovascular Digital Twins}

Cardiovascular digital twins vary in fidelity depending on clinical objectives
and computational constraints. Organ-scale twins are most common and focus on
the heart, vasculature, or their interaction. Examples include patient-specific
electromechanical heart models and haemodynamic vascular twins that estimate
flow, pressure, and wall stress. These models balance physiological
interpretability with computational feasibility~\cite{arzani2022machine,viola2023gpu}.

Multi-scale twins extend this framework by coupling processes across cellular,
tissue, organ, and systemic levels. In cardiovascular modelling, this may
involve linking cellular electrophysiology with myocardial mechanics or
integrating local haemodynamics with global circulation models. Although
multi-scale coupling captures emergent phenomena such as arrhythmogenesis and
remodelling, it introduces challenges in parameter identifiability and
computational burden~\cite{li2024solving,kerckhoffs2006computational}.

A further dimension arises from multi-physics integration. Cardiovascular
function reflects tightly coupled electrophysiology, solid mechanics, and
fluid dynamics. Digital twins increasingly incorporate electromechanical
simulations and fluid-structure interaction models to improve physiological
fidelity. However, higher fidelity typically comes at the cost of scalability
and real-time deployment~\cite{xu2025cardiac,tesan2025thermodynamics}.

\subsection{Clinical Data Sources and Modalities}

Cardiovascular digital twins rely on heterogeneous clinical data. Imaging
modalities such as magnetic resonance imaging, computed tomography, and
echocardiography provide structural and functional information and underpin
many patient-specific models~\cite{canino2025artificial,bewig2025cardiovascular}.

Physiological measurements, including blood pressure, flow velocity, and
electrocardiographic signals, complement imaging data and support parameter
calibration. Importantly, certain digital twin applications, such as
continuous non-invasive blood pressure monitoring or haemodynamic surveillance
using 0D models, may operate without imaging inputs, relying entirely on
physiological waveforms acquired through wearable or bedside
sensors~\cite{zhang2020personalized,gray2018patient}. Wearable sensors extend
this paradigm by enabling longitudinal monitoring of heart rate, rhythm, and
activity outside clinical environments, thereby facilitating continuous
updating and real-time adaptation~\cite{rudnicka2024cardiac,sel2024building}.

Electronic health records contribute contextual information, including
comorbidities, medications, and laboratory results. When integrated with
mechanistic and learning-based components, such data support personalisation
and population-to-individual translation. However, data heterogeneity,
missingness, and interoperability remain significant barriers~\cite{meijer2023digital,strocchi2025cardiac}.

Emerging efforts explore the incorporation of molecular and omics data to
enhance biological specificity. Although promising, integration of genomics
and proteomics substantially increases model complexity and raises challenges
related to interpretability and validation, positioning this area as an active
research frontier rather than an established clinical
capability~\cite{huang2024application,sel2024building}.

\subsection{Comparative Overview of Modelling Paradigms}
 
The modelling paradigms reviewed in this paper, spanning 0D lumped-parameter
models, 1D haemodynamic formulations, 3D CFD and electromechanical solvers,
machine learning, PINNs, GNNs, and hybrid frameworks, differ substantially in
their computational cost, data requirements, interpretability, scalability,
uncertainty quantification maturity, and approximate clinical readiness.
Table~\ref{tab:paradigm_comparison} provides a structured qualitative
comparison across these dimensions, based on representative published studies.
This summary is intended to orient readers before the detailed paradigm-by-paradigm
treatment in subsequent sections, and to facilitate direct comparison of the
trade-offs each approach entails for digital twin deployment.

\begin{landscape}
\begin{table}[!t]
\caption{Comparative summary of cardiovascular digital twin modelling
paradigms across key dimensions. Ratings are qualitative and based on
representative published studies; individual implementations may vary.
Cost: computational cost at inference/deployment. Data req.: data
requirements for personalisation. Interp.: interpretability.
UQ: uncertainty quantification maturity. CR: approximate clinical
readiness.}\label{tab:paradigm_comparison}
\begin{tabular*}{\linewidth}{@{\extracolsep\fill}
  l
  p{2.2cm}
  p{2.2cm}
  p{1.8cm}
  p{2.2cm}
  p{2.0cm}
  p{2.2cm}}
\toprule
Paradigm & Comp.\ Cost & Data Req. & Interp. & Scalability & UQ Maturity & Clin.\ Readiness \\
\midrule
0D Lumped-parameter  & Very low        & Low             & High           & High           & Moderate      & High           \\
1D Haemodynamic      & Low             & Low--Moderate   & High           & High           & Moderate      & Moderate--High \\
3D CFD / FSI         & Very high       & High            & High           & Low            & Low--Moderate & Low            \\
Electromechanical    & Very high       & Very high       & High           & Very low       & Low           & Low            \\
ML / Deep learning   & Low (inference) & High (training) & Low            & High           & Low           & Moderate       \\
PINNs                & Moderate        & Low--Moderate   & Moderate       & Low--Moderate  & Low           & Low--Moderate  \\
GNNs                 & Low--Moderate   & Moderate        & Moderate       & Moderate--High & Low           & Low--Moderate  \\
Hybrid frameworks    & Variable        & Variable        & Moderate--High & Moderate       & Moderate      & Moderate       \\
\botrule
\end{tabular*}
\end{table}
\end{landscape}

\section{Biophysical and Mechanistic Modelling Approaches}
\label{sec:biophysical}

Biophysical modelling forms the mechanistic foundation of many cardiovascular digital twins. These approaches derive system behaviour from physical principles governing haemodynamics, tissue mechanics, and electrophysiology. They offer interpretability and physiological consistency but vary substantially in fidelity, computational cost, and scalability.

\subsection{Lumped-Parameter and One-Dimensional Models}

\begin{figure}[htbp]
    \centering
    \includegraphics[width=\linewidth]{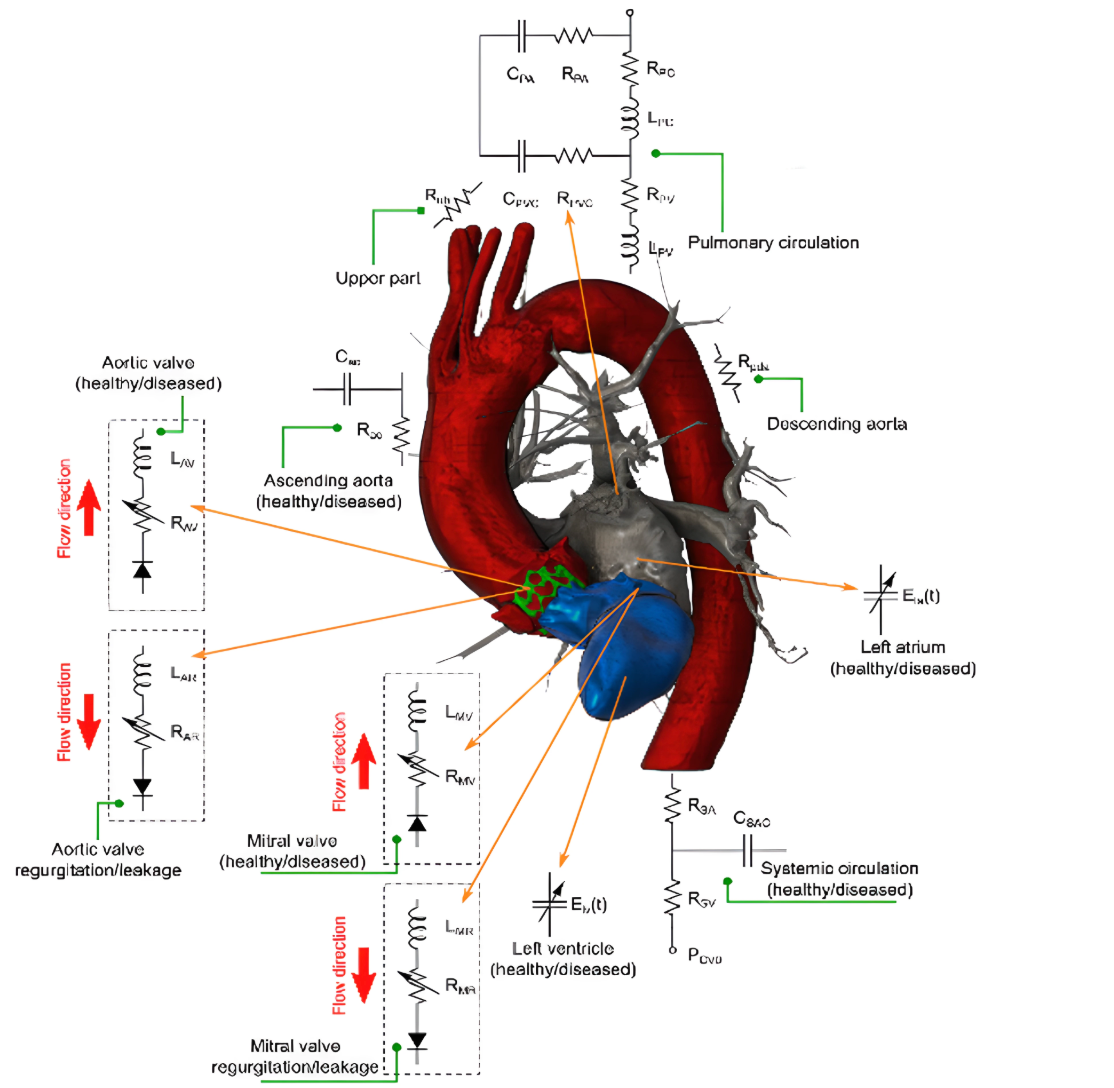}
    \caption{Lumped-parameter cardiovascular model coupled with 3D heart anatomy. Electrical analogue elements represent systemic and pulmonary circulations, cardiac chambers, and valve dynamics.}
    \label{fig:cardio_model}
\end{figure}

Lumped-parameter (zero-dimensional, 0D) models, illustrated in Figure~\ref{fig:cardio_model}, represent the cardiovascular system as a network of interconnected compartments characterised by spatially averaged pressure, flow, and volume variables. Originating from analogies with electrical circuits, these models employ resistive, capacitive, and inertial elements to reproduce global hemodynamic behaviour. Classical Windkessel formulations remain fundamental in this context and continue to provide the basis for closed-loop circulation models that incorporate cardiac chambers and valves \cite{mou2015exploring,naik2017mathematical,shi2013lumped}.

Modern 0D models are widely used to study ventricular–arterial coupling, cardiac output regulation, and blood pressure dynamics, and they are particularly attractive for real-time simulation and integration into digital twin pipelines \cite{zhang2020personalized,mynard2015one,gray2018patient}. Their computational efficiency makes them suitable as system-level backbones or boundary condition generators for higher-fidelity models. However, spatial averaging limits their ability to capture wave propagation and localised flow disturbances, constraining their predictive resolution in anatomically complex diseases.

At the subcellular and cellular level, ordinary differential equation (ODE)-based ionic models, such as the Hodgkin-Huxley formalism and its cardiovascular successors including the Luo-Rudy and ten~Tusscher models, describe the dynamics of transmembrane ion currents and action potential generation. Although these models are rarely deployed in isolation as clinical digital twins for human patients, they constitute the foundational computational unit for whole-heart electrophysiological simulations and are central to experimental digital twins used in drug safety screening and ion channel pharmacology. In this context, in-silico trials employing virtual populations of cellular models offer a compelling complement to preclinical
drug development pipelines, providing mechanistic insight into proarrhythmic risk that is difficult to obtain from animal models alone~\cite{mirams2016uncertainty,trayanova2011whole}.

\begin{figure}[htbp]
    \centering
    \includegraphics[width=0.8\linewidth]{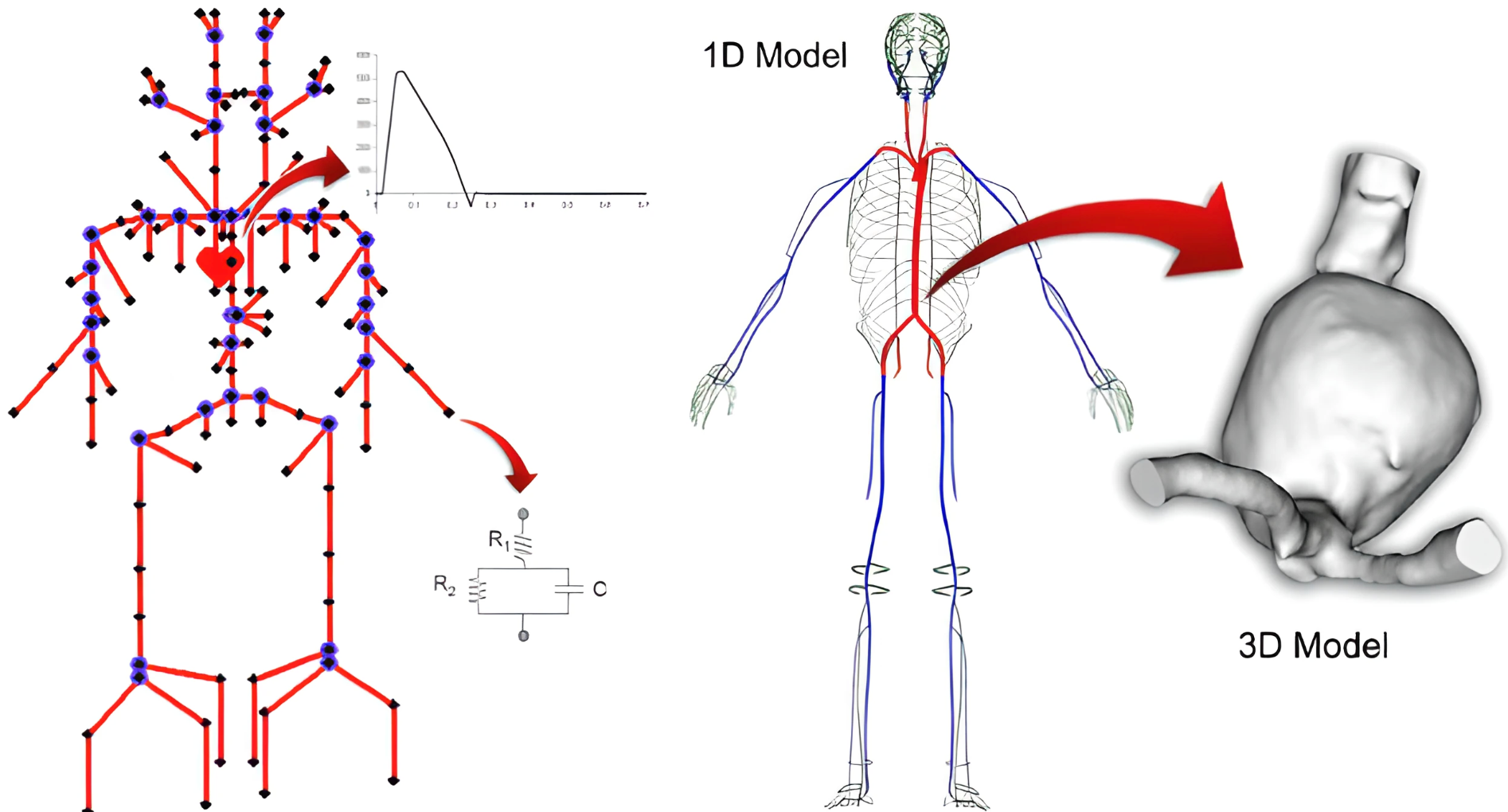}
    \caption{Multiscale cardiovascular modelling. A one-dimensional arterial network captures systemic wave propagation, while three-dimensional models resolve local flow structures. Reproduced with permission from Larrabide et al.~\cite{larrabide2012hemolab}.}
    \label{fig:1D_3D_model}
\end{figure}

One-dimensional (1D) haemodynamic models, as illustrated in Figure~\ref{fig:1D_3D_model}, extend the 0D framework by resolving pressure
and flow along vessel centrelines. Derived from cross-sectionally averaged
Navier-Stokes equations, these models explicitly capture pulse wave propagation, reflection, and arterial stiffness effects~\cite{mynard2015one,shi2011review}..

Arterial tree representations based on 1D models are widely used to study
systemic and pulmonary circulation, assess arterial stiffness, and quantify
wave reflections related to ageing and cardiovascular disease. Compared with
0D models, 1D formulations provide much higher physiological fidelity while
remaining computationally efficient for large-scale parameter studies and
patient-specific simulations~\cite{shi2011review}, making them well suited to
digital twin architectures that require repeated model updates and forward
simulations. They are frequently used to infer central pressure from peripheral
measurements or to provide boundary conditions for localised 3D
simulations~\cite{cai2024lumped}. Nevertheless, assumptions regarding flow
symmetry and simplified vessel geometry limit accuracy in regions with strong
three-dimensional flow complexity.

\subsection{Three-Dimensional Hemodynamic and Electromechanical Models}

Three-dimensional (3D) computational models offer the highest spatial resolution in cardiovascular modelling. Computational fluid dynamics (CFD) simulations solve the Navier–Stokes equations within patient-specific geometries reconstructed from imaging, enabling quantification of wall shear stress, pressure gradients, and complex flow patterns associated with atherosclerosis and thrombosis \cite{frangi2002three,colebank2024guidelines}.

To account for vessel and myocardial deformation, fluid–structure interaction (FSI) frameworks couple haemodynamics with solid mechanics descriptions of tissue behaviour \cite{tesan2025thermodynamics,mynard2015one}. These models are particularly relevant in valve disease, aneurysm assessment, and intervention planning. However, increased fidelity is accompanied by substantial computational costs and demanding data requirements.

\begin{figure}[htbp]
    \centering
    \includegraphics[width=0.8\linewidth]{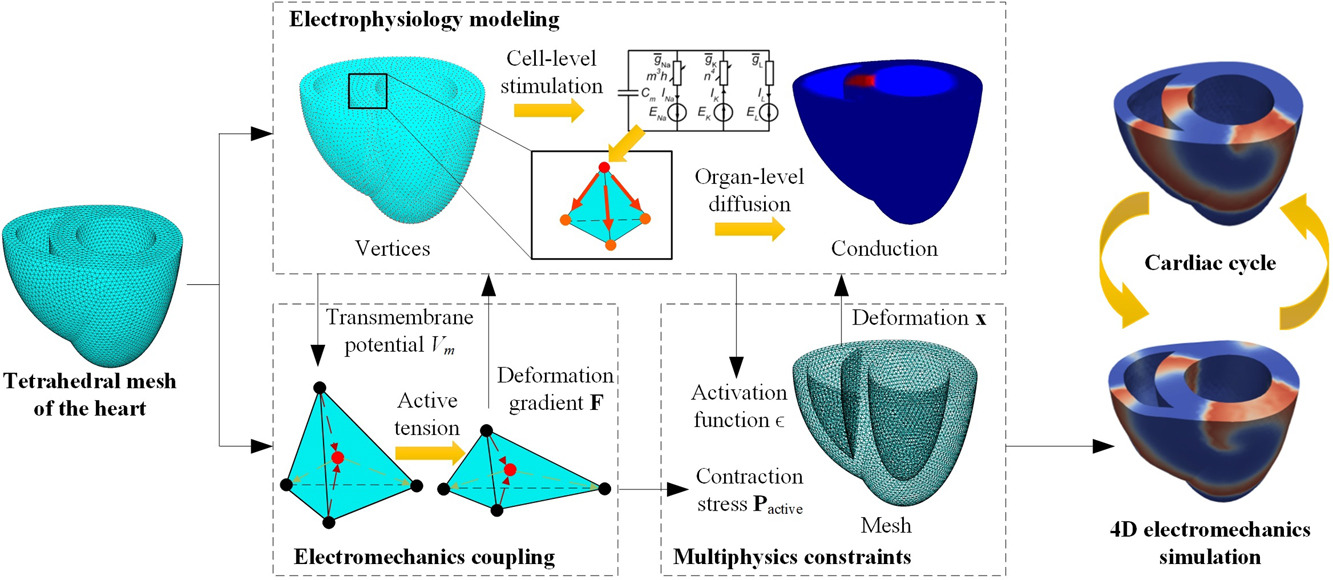}
    \caption{Multiscale cardiac electrophysiology and electromechanics framework linking cellular excitation, tissue conduction, myocardial contraction, and organ-scale deformation. Reproduced with permission from Chen et al.~\cite{chen2024coupling}.}
    \label{fig:electromech_framework}
\end{figure}

Electrophysiology and electromechanics represent the most sophisticated layer of mechanistic modelling. Electrophysiological formulations describe the generation and propagation of electrical activation using cellular ionic models embedded within monodomain or bidomain tissue equations. Whole-heart implementations have reproduced arrhythmias, guided ablation strategies, and predicted defibrillation outcomes in patient-specific settings \cite{trayanova2011whole,smith2011euheart}.
An overview of a multiscale cardiac electrophysiology and electromechanics framework is shown in Figure~\ref{fig:electromech_framework}.

Electromechanical models extend this framework by coupling excitation–contraction mechanisms to nonlinear tissue mechanics, enabling the simulation of ventricular filling, ejection, and wall stress under physiological and pathological conditions \cite{kerckhoffs2006computational,sun2014computational}. Coupling with circulation models further enables the prediction of pressure–volume dynamics and ventricular–arterial interaction \cite{bucelli2023mathematical,mynard2015one}. These integrated models approach comprehensive whole-heart digital twins capable of linking electrical, mechanical, and hemodynamic dysfunction.

Despite their physiological richness, 3D and electromechanical models require extensive preprocessing, high-resolution imaging, fibre architecture reconstruction, and parameter calibration. Simulation times may extend to hours or days per scenario, limiting routine clinical deployment.

\subsection{Personalisation and Uncertainty}
\label{sec:personalisation}

Personalisation is central to digital twin deployment but remains challenging for high-fidelity mechanistic models. Parameter estimation is typically formulated as an inverse problem, using optimisation, adjoint methods, or Bayesian inference to infer model parameters from clinical data \cite{gul2016mathematical,mirams2016uncertainty}. Data assimilation strategies, including Kalman filtering and variational approaches, enable dynamic updating as new measurements become available \cite{duanmu2019one,sun2014computational}.

Cardiovascular data are typically sparse relative to the dimensionality of the underlying models, rendering personalisation a fundamentally data-limited inverse problem. In practice, only a small number of observable outputs, such as pressure waveforms, volumetric flow rates, or wall motion measurements, are available to constrain model parameters that may number in the hundreds. This observational sparsity produces ill-conditioned estimation problems, non-unique solutions, and large posterior uncertainties that must be propagated through the model to obtain reliable predictive intervals. Surrogate-based uncertainty propagation, while computationally tractable, introduces additional approximation error that can inflate uncertainty bounds in a manner difficult to quantify without extensive
benchmarking~\cite{mirams2016uncertainty,colebank2024guidelines}.

A further underappreciated source of uncertainty is model-form discrepancy: the systematic deviation between model predictions and reality attributable to
idealised constitutive laws, simplified geometry, or unmodelled physiological
processes. This type of structural error is not reducible by additional data
collection alone and requires explicit treatment through model discrepancy
terms or ensemble approaches~\cite{rodero2023advancing,garber2022critical}.
Reference to the ASME V\&V 40 standard (Assessing Credibility of Computational Modelling through Verification and Validation: Application to Medical Devices) provides a formal risk-based framework for credibility assessment of computational models, with principles that are applicable to cardiovascular digital twins~\cite{colebank2024guidelines}.

A foundational distinction in uncertainty characterisation is between
\emph{aleatory} uncertainty, irreducible variability arising from biological
stochasticity, inter-patient heterogeneity, and measurement noise, and
\emph{epistemic} uncertainty, which reflects incomplete knowledge of model
parameters, structure, or boundary conditions and is in principle reducible
with additional data. Both types are present simultaneously in cardiovascular
digital twins, and failure to distinguish them can lead to miscalibrated
predictions and inappropriate clinical confidence.

\emph{Structural identifiability} refers to whether a unique parameter set can in principle be recovered from perfect, noise-free data given the model structure, a property that can be assessed analytically for ODE systems. \emph{Practical identifiability}, by contrast, concerns whether parameters can be reliably estimated from the finite, noisy data available in clinical practice. Many cardiovascular models exhibit structural identifiability but practical non-identifiability, meaning that while a unique solution exists theoretically, it cannot be recovered robustly from real measurements. Profile likelihood analysis, Bayesian posterior geometry inspection, and sensitivity-based methods are standard tools for diagnosing practical identifiability issues~\cite{mirams2016uncertainty,alonso2025biophysical}.

Parameter identifiability poses a persistent limitation more broadly. Many
models contain high-dimensional parameter spaces that are only partially
constrained by available data, leading to non-unique solutions and prediction
uncertainty~\cite{alonso2025biophysical}. Sensitivity to small parameter
variations is particularly pronounced in electrophysiological systems, where
minor changes in ionic conductances may trigger qualitatively different
rhythms~\cite{mirams2016uncertainty}. These challenges underscore the
importance of uncertainty quantification and model reduction strategies.

\subsection{Limitations of Purely Mechanistic Digital Twins}

While mechanistic models provide interpretability and adherence to physical laws, their limitations become evident when considered as standalone digital twins. High computational costs restrict real-time application, and extensive data requirements reduce scalability \cite{xu2025cardiac,alonso2025biophysical}. Model complexity introduces multiple sources of uncertainty, including geometric reconstruction, boundary condition specification, and the choice of constitutive parameters \cite{rodero2023advancing,garber2022critical}.

Moreover, purely biophysical approaches struggle to accommodate sparse, noisy, or heterogeneous clinical datasets. As digital twins move toward continuous updating and longitudinal monitoring, the need for rapid inference and scalable computation becomes increasingly critical.

These limitations have motivated the integration of reduced-order modelling, data-driven surrogates, and physics-informed learning frameworks that aim to preserve mechanistic consistency while improving computational efficiency and robustness \cite{wang20213d,sel2024building}. The following sections examine these emerging paradigms and their role in bridging interpretability and scalability in cardiovascular digital twins.

%=================================================================
\section{Data-Driven Models for Cardiovascular Digital Twins}
\label{sec:data_driven}

Data-driven modelling has become increasingly prominent in cardiovascular research, driven by expanding clinical datasets and advances in machine learning (ML). Unlike mechanistic models that encode physiological laws explicitly, ML models learn mappings directly from data. Within cardiovascular digital twins, these approaches are primarily used to enhance scalability, accelerate computation, and infer latent physiological quantities that are difficult to measure directly.

\subsection{Machine Learning for Physiological Inference}

Early ML applications in cardiology focused on classification and regression tasks, including risk stratification, disease detection, and outcome prediction from structured clinical and imaging-derived variables \cite{dinh2019data,van2025individual}. Although these approaches achieved strong predictive performance at the population level, they offered limited mechanistic insight and were initially deployed as decision-support tools rather than as core components of physiological digital twins.

More recent work integrates ML into modelling pipelines to approximate constitutive relations, estimate model parameters, and emulate computationally expensive solvers. In this context, ML acts as a modelling accelerator rather than a replacement for physics-based descriptions \cite{henglin2017machine,arzani2021data}. Within digital twin architectures, learning-based components are increasingly used to map sparse clinical observations to latent states, such as estimating central blood pressure from peripheral measurements or inferring material properties from imaging data \cite{regazzoni2021combining,bauer2023data}. These capabilities are particularly valuable for longitudinal updating, where repeated invasive measurements are impractical.

\subsection{Deep Learning for Imaging and Time Series}

Deep learning (DL) has transformed cardiovascular modelling, especially in high-dimensional settings. Convolutional neural networks (CNNs) are widely used for segmentation, registration, and feature extraction in MRI, CT, and echocardiography \cite{henglin2017machine,coorey2022health}. Automated image processing is foundational for digital twin construction, enabling scalable reconstruction of patient-specific geometries and functional descriptors.

Beyond preprocessing, CNN-based models have been applied to predict ventricular volumes, ejection fraction, and even flow-related quantities directly from imaging data \cite{gandin2021interpretability,arzani2022machine}. While often accurate within training distributions, purely data-driven predictions may lack robustness under extrapolation, limiting reliability in heterogeneous clinical environments.

For temporal data such as electrocardiograms, pressure waveforms, and wearable sensor streams, recurrent neural networks, long short-term memory networks, and Transformer architectures capture temporal dependencies and long-range correlations \cite{dritsas2023efficient,kissi2025data}. In digital twin contexts, these models support state estimation and short-term forecasting between clinical encounters. However, concerns regarding interpretability, calibration, and uncertainty remain significant barriers to clinical trust \cite{henglin2017machine,barzegar2025predictive}.

\subsection{Surrogate and Reduced-Order Modelling}

A central role of ML in cardiovascular digital twins lies in surrogate modelling. Here, learning-based models emulate the input–output behaviour of high-fidelity simulations such as CFD, fluid–structure interaction, or electromechanical solvers. Once trained, surrogates can generate predictions orders of magnitude faster than full numerical simulations, enabling real-time parameter sweeps, optimisation, and uncertainty analysis.

Reduced-order models (ROMs), constructed using proper orthogonal decomposition or reduced basis techniques, further decrease computational burden while retaining dominant dynamical features. Hybrid approaches combine ROMs with ML to enhance generalisation and calibration \cite{morid2023time,shameer2018machine}. In digital twin workflows, such surrogates accelerate parameter estimation and probabilistic inference, facilitating patient-specific exploration of large parameter spaces \cite{arzani2021data,regazzoni2021combining}.

Nevertheless, surrogate performance depends strongly on training data coverage. Models trained within narrow parameter ranges may fail under extrapolation, raising safety concerns in clinical deployment \cite{prabhu2023data,gerdroodbary2025predictive}. These limitations highlight the importance of validation and the incorporation of physical constraints.

\subsection{Advantages and Limitations}

Data-driven approaches offer compelling advantages for cardiovascular digital twins. They are computationally efficient, scalable across large populations, and capable of integrating heterogeneous data sources. These properties support personalisation, longitudinal monitoring, and population-to-individual translation \cite{bauer2023data,barzegar2025predictive}. In particular, ML alleviates computational bottlenecks associated with high-fidelity biophysical simulations.

However, several limitations constrain purely data-driven twins. Generalisation beyond the training distribution remains a central challenge, especially in the presence of dataset shift across institutions, imaging protocols, or disease phenotypes \cite{henglin2017machine,dinh2019data}. Interpretability is another barrier: unlike mechanistic models grounded in physiology, ML systems often function as black boxes. While explainable AI techniques aim to address this issue, consensus on robust interpretability standards in cardiovascular medicine is still evolving \cite{arzani2021data,dozen2020image}.

Uncertainty quantification is frequently underdeveloped. Deterministic predictions may convey unwarranted confidence in safety-critical contexts such as intervention planning. Increasingly, the literature emphasises probabilistic modelling and the integration of physical constraints to improve robustness and trustworthiness \cite{barzegar2025predictive,coorey2022health}.

Taken together, data-driven models play a critical but complementary role in cardiovascular digital twins. Their scalability and speed address key limitations of purely mechanistic approaches, yet challenges in extrapolation, interpretability, and uncertainty limit standalone deployment. These considerations have directly motivated the development of physics-informed neural networks and graph-based learning methods, which seek to integrate physical consistency with data efficiency and are discussed in the following section.

%=================================================================
\section{Physics-Informed Neural Networks}
\label{sec:pinns}

Physics-informed neural networks (PINNs) represent a central development in scientific machine learning, embedding governing physical laws directly into neural network training. By incorporating differential equations and conservation principles into the loss function, PINNs constrain learned solutions to remain consistent with known physics. This approach contrasts with purely data-driven models that rely exclusively on input–output mappings and may violate fundamental physical relationships.

\subsection{Principles of Physics-Informed Learning}
\label{sec:pinn_principles}

\begin{figure}[htbp]
\centering
\includegraphics[width=0.8\linewidth]{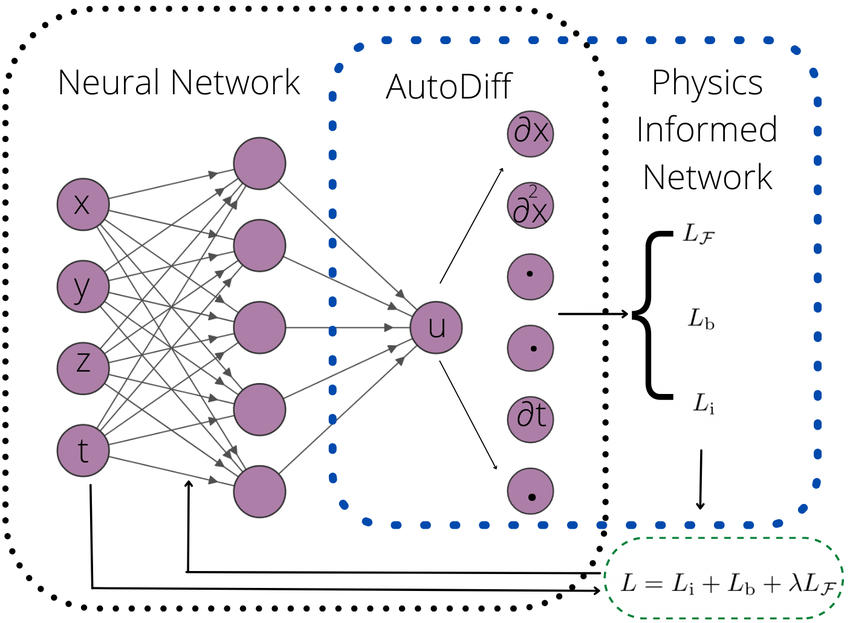}
\caption{Schematic of a Physics-Informed Neural Network (PINN). Inputs
$(x,y,z,t)$, and optionally patient-specific or physical parameters
$\boldsymbol{\theta}$ (e.g.\ viscosity, boundary resistances, material
constants) in parametric formulations, are fed into a neural network,
producing the predicted field $u$. Automatic differentiation (AutoDiff)
computes derivatives required to evaluate the PDE residuals. The losses
$\mathcal{L}_{i}$, $\mathcal{L}_{b}$, and $\mathcal{L}_{\mathcal{F}}$
correspond to data-fit, boundary-condition, and PDE residual terms,
respectively, combined as
$\mathcal{L} = \mathcal{L}_{i} + \mathcal{L}_{b} + \lambda\,\mathcal{L}_{\mathcal{F}}$,
where $\lambda>0$ is a weighting hyperparameter that balances the relative
contribution of the physics residual against data and boundary losses. This
structure enables PINNs to approximate solutions of ODEs or PDEs while
embedding physical constraints.}
\label{fig:pinn_schematic}
\end{figure}

Physics-informed neural networks (PINNs) approximate the solution of ordinary
or partial differential equations using neural networks, while enforcing
governing equations at collocation points through automatic differentiation
(Figure~\ref{fig:pinn_schematic})~\cite{raissi2019physics,karniadakis2021physics}.
In this schematic, inputs $(x,y,z,t)$ are passed through a neural network to
predict $u$, and AutoDiff computes the necessary derivatives to evaluate the
PDE residuals. In parametric PINN formulations, the network may additionally
receive physical or patient-specific parameters $\boldsymbol{\theta}$ (such as
fluid viscosity, boundary resistances, or material constants) as inputs alongside the spatiotemporal coordinates, enabling rapid evaluation across a parameter space without retraining, a feature particularly relevant to digital twin personalisation~\cite{wang2023expert}.

The PDE residual loss for cardiovascular haemodynamics can be expressed
concretely in terms of the incompressible Navier-Stokes equations as:
\begin{equation}
\mathcal{L}_{\mathcal{F}} = \frac{1}{N_c}\sum_{j=1}^{N_c}
\Bigl[\bigl\|\rho(\mathbf{u}\cdot\nabla)\mathbf{u}
  + \nabla p - \mu\nabla^{2}\mathbf{u}\bigr\|^{2}
+ \|\nabla\cdot\mathbf{u}\|^{2}\Bigr]_{\mathbf{x}_{j}},
\label{eq:pinn_ns_loss}
\end{equation}
where $\mathbf{u}$ is the velocity field, $p$ is pressure, $\rho$ is density,
$\mu$ is dynamic viscosity, and $N_c$ is the number of collocation points. The
total loss is then:
\begin{equation}
\mathcal{L} = \mathcal{L}_{i} + \mathcal{L}_{b}
  + \lambda\,\mathcal{L}_{\mathcal{F}},
\label{eq:pinn_total_loss}
\end{equation}
where $\lambda>0$ is the physics loss weighting hyperparameter. The choice of
$\lambda$ is problem-dependent and has significant impact on training
convergence and solution accuracy; adaptive schemes that adjust $\lambda$
during training have been proposed to mitigate this
sensitivity~\cite{wang2023expert,hao2022physics}.

This unified formulation enables forward simulation, parameter identification,
and data assimilation within a single framework. By embedding physical
constraints, PINNs introduce implicit regularisation that improves
generalisation in settings characterised by sparse, noisy, or indirect
biomedical measurements~\cite{arzani2021uncovering,farea2024understanding}.

Importantly, PINNs are not intended to replace classical solvers in
high-accuracy numerical analysis. Instead, they trade strict convergence
guarantees for flexibility, data efficiency, and seamless integration with
learning-based pipelines~\cite{huang2022applications,sharma2023review}. These
characteristics align closely with the requirements of cardiovascular digital
twins, where rapid updating and personalisation are often prioritised over
high-order numerical precision.

\subsection{PINNs for Cardiovascular Hemodynamics}

Cardiovascular haemodynamics has emerged as a leading application area for PINNs, given the well-defined Navier–Stokes framework and the computational cost of conventional CFD. PINNs have been applied to reconstruct pressure and velocity fields from sparse measurements, including limited 4D-flow MRI data and catheter-based recordings \cite{conti2024multi,li2024solving}. By enforcing conservation of mass and momentum, these models infer unmeasured quantities and recover physically consistent flow fields even when boundary conditions are uncertain.

PINN-based surrogates have been proposed to estimate clinically relevant metrics such as wall shear stress and pressure drops across stenoses with substantially reduced computational cost compared to full CFD \cite{taebi2022deep,arzani2022machine}. Unlike purely data-driven surrogates, the embedded physics improves robustness under moderate extrapolation by constraining predictions to physiologically admissible regimes.

Beyond forward simulation, PINNs are particularly attractive for inverse problems central to digital twin personalisation. Studies demonstrate their ability to estimate boundary conditions, flow resistance, and material parameters directly from sparse clinical data \cite{wang2023expert,barzegar2025predictive}. Such capabilities support continuous updating of vascular digital twins without repeated high-fidelity simulations.

A significant practical bottleneck in cardiovascular PINN applications is the
treatment of patient-specific geometry. Unlike mesh-based finite element
solvers that discretise arbitrary domains, PINNs represent the computational
domain implicitly, typically through coordinate transformations, signed
distance functions, or level-set representations. Patient-specific vascular
geometries, derived from segmented imaging data, are often complex, branching,
and multi-connected, posing substantial challenges for collocation point
sampling, boundary condition enforcement, and training convergence. Domain
decomposition strategies, wherein separate networks are trained on geometric
subdomains with interface continuity constraints, have been proposed as a
mitigation, but these increase implementation complexity and training
cost~\cite{wang2023expert,hao2022physics}. The sensitivity of PINN accuracy to
geometric fidelity and domain representation remains an active and largely open
research problem.

Scalability remains a challenge more broadly. Training instability, sensitivity
to loss weighting, and performance degradation in complex three-dimensional
geometries limit current implementations, many of which focus on reduced-order
or simplified domains.

\subsection{PINNs for Electrophysiology and Mechanics}
\label{sec:pinns_ep}

Physics-informed neural networks (PINNs) are being increasingly investigated in the context of cardiac electrophysiology and electromechanics, domains in which conventional numerical solvers are computationally expensive and parameter estimation is challenging. In electrophysiology, physics-informed strategies have been employed for both monodomain and bidomain formulations to reconstruct spatiotemporal activation patterns and to infer tissue conduction properties from sparse electrocardiographic (ECG) or intracardiac measurements \cite{olakorede2022physics,sarabian2022physics}. The explicit incorporation of the governing equations attenuates spurious, non-physiological dynamics that are frequently observed in purely data-driven, black-box models and promotes physiologically consistent parameter inference.

In cardiac mechanics, PINN-based frameworks have been proposed to approximate the governing equations of nonlinear elasticity and excitation–contraction coupling, thereby enabling the estimation of myocardial material parameters and active stress components from limited imaging-derived observations \cite{panneerselvam2026toward,lydon2025physics}. These advances are particularly pertinent to the construction of patient-specific cardiac digital twins, for which many mechanical parameters are not directly accessible through in vivo measurement.

As in haemodynamics, the representation of patient-specific cardiac geometry
within PINN frameworks is a significant bottleneck. The three-dimensional,
anisotropic, and fibre-structured myocardium poses greater geometric complexity
than simple vascular geometries, making accurate collocation sampling and
boundary condition enforcement particularly demanding.

Notwithstanding these promising developments, the robust coupling of stiff electrophysiological dynamics with highly nonlinear tissue mechanics remains numerically challenging. Analogous to the situation in cardiovascular haemodynamics, achieving scalable and computationally tractable PINN formulations for fully three-dimensional, multi-physics heart models continues to represent a major open research problem.

\subsection{Hybrid Biophysical–PINN Architectures}

The most compelling role of PINNs in cardiovascular digital twins lies in hybrid architectures. Rather than replacing mechanistic models, PINNs can function as surrogate solvers, correction operators, or parameter estimators embedded within larger biophysical pipelines \cite{regazzoni2021combining,arzani2021data}. In such systems, reduced-order or compartmental models provide global structure, while PINNs infer local dynamics or compensate for modelling discrepancies.

Hybrid approaches are particularly suited to real-time updating. By learning unknown or patient-specific components while preserving known physical structure, PINNs enable adaptive digital twins that integrate streaming data without repeated full-scale numerical simulation \cite{herrero2022ep,meng2025physics}. This positioning situates PINNs between purely mechanistic solvers and black-box ML models, balancing interpretability, computational efficiency, and data integration \cite{jeske2020digital,kissas2023towards}.

\subsection{Limitations and Open Challenges}
\label{sec:pinn_limits}

Despite their conceptual appeal, several limitations restrict widespread clinical translation. 
Several well-documented training pathologies limit PINN performance in cardiovascular settings. \emph{Spectral bias}, the tendency of neural
networks to preferentially learn low-frequency components of the solution, poses particular difficulties for cardiac electrophysiology, where action potential
wavefronts involve sharp spatial gradients and rapid temporal transitions that
require high-frequency representation. \emph{Stiff PDE systems}, characteristic of ionic models where gating variables evolve on timescales orders of magnitude
faster than the action potential, exacerbate training instability by inducing
poorly conditioned loss landscapes. \emph{Gradient pathologies} including
vanishing and exploding gradients across deep networks are compounded in multi-physics settings where loss terms from PDE residuals, boundary conditions, and observational data may differ by several orders of magnitude. \emph{Loss imbalancing}, the failure of naive gradient descent to appropriately weight competing loss components, frequently causes the optimiser to minimise one term at the expense of others, producing solutions that satisfy the PDE but violate data constraints or vice versa. Adaptive loss weighting schemes, learning rate annealing, and residual-based adaptive sampling strategies have been proposed, but no universal solution exists~\cite{wang2023expert,hao2022physics}. These challenges are most acute in three-dimensional, patient-specific geometries where collocation point requirements scale unfavourably, and in coupled electromechanical systems where feedback between stiff ionic dynamics and nonlinear large-deformation mechanics creates further numerical difficulties.

Scalability is a central challenge. While PINNs perform well in low-dimensional settings, performance may degrade for complex vascular networks or whole-heart electromechanics. Techniques such as domain decomposition, adaptive sampling, and multi-network architectures are under active investigation \cite{wang2023expert,panneerselvam2026toward}.

Uncertainty quantification is another unresolved issue. Most cardiovascular PINN implementations provide deterministic predictions, despite measurement noise and model discrepancy. Bayesian PINNs and ensemble strategies have been proposed, but systematic integration into digital twin frameworks remains limited \cite{barzegar2025predictive,farea2024understanding}.

Ultimately, clinical translation necessitates stringent validation and systematic benchmarking. Although PINNs embed governing physical principles, their outputs remain contingent on learned latent representations that must be rigorously compared against established numerical solvers and high-fidelity clinical ground truth. Ongoing progress in numerical stability, computational scalability, and uncertainty-aware learning will be pivotal in determining whether PINN-based cardiovascular digital twins can evolve from experimental research prototypes into robust tools for routine clinical practice.

In parallel, the distinct but complementary advantages and drawbacks of PINNs have stimulated interest in graph-based learning methodologies that explicitly exploit the networked topology of the cardiovascular system, as elaborated in the subsequent section.

%=================================================================
\section{Graph Neural Networks for Cardiovascular Digital Twins}
\label{sec:gnns}

\subsection{Graph Representations of the Cardiovascular System}

\begin{figure}[H]
\centering
\includegraphics[width=0.8\linewidth]{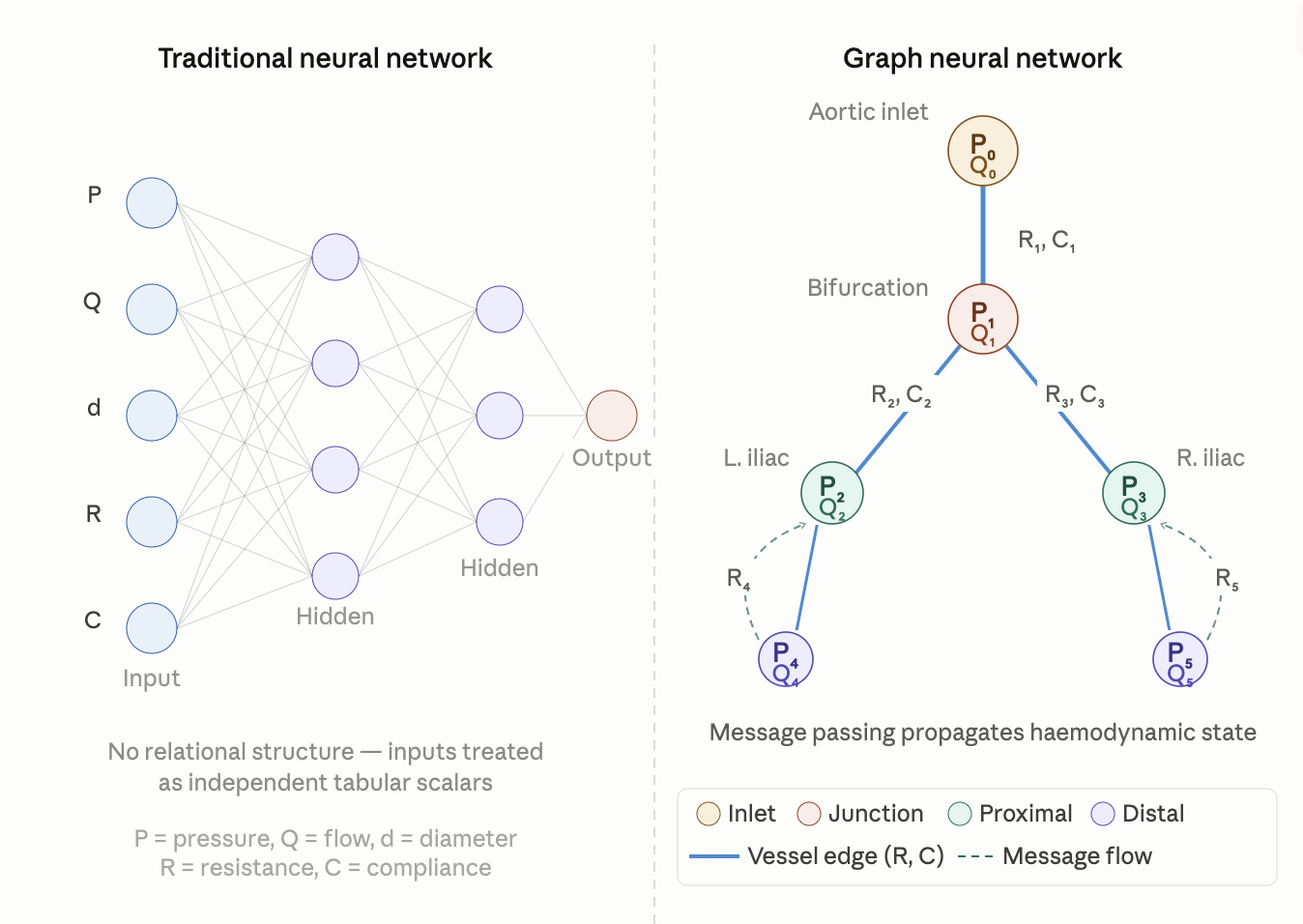}

\caption{Comparison of traditional neural networks (left) and graph neural
networks (right). GNNs propagate information along edges to exploit relational
structure. In cardiovascular applications, GNN nodes correspond to vessel
segments or bifurcations with features such as pressure and flow, while edges
encode anatomical connectivity, resistance, and compliance, structure absent
in conventional neural networks operating on tabular haemodynamic inputs.}
\label{fig:tnn_vs_gnn}
\end{figure}

Graphs model objects (nodes) and their relationships (edges). Graph neural networks (GNNs) are tailored for networked data, where the state of a node depends on its neighbours. The cardiovascular system is naturally graph-structured: vessels and functional compartments are connected in branching networks, and measurements such as pressure, flow, and diameter exhibit relational dependencies \cite{barbiero2021graph}. This network bias makes GNNs suitable for scalable digital twin models, particularly when full 3D simulations are computationally prohibitive (Figure~\ref{fig:tnn_vs_gnn}).

\begin{figure}[H]
\centering
\includegraphics[width=0.65\linewidth]{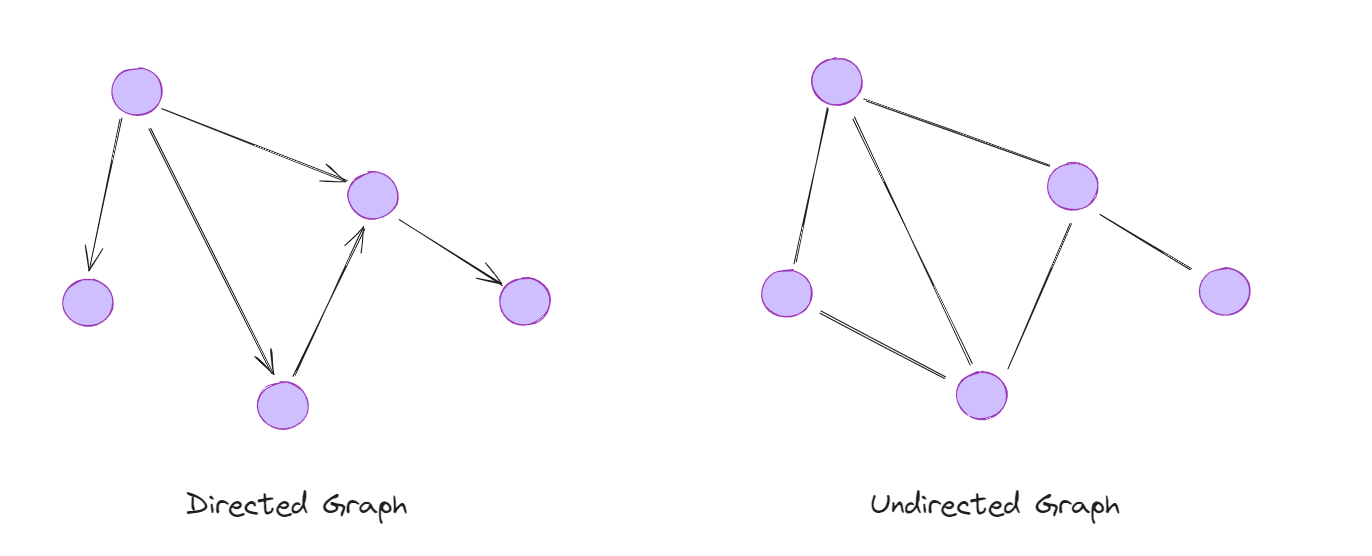}
\caption{Directed and undirected graph representations. Directed edges capture
flow causality in arterial trees; undirected edges represent bidirectional or
symmetric relations such as structural adjacency in myocardial tissue.}
\label{fig:directed_undirected}
\end{figure}

Directed graphs are appropriate for arterial trees, encoding flow direction and causality, while undirected graphs represent symmetric interactions or structural adjacency (Figure~\ref{fig:directed_undirected}). Nodes may correspond to bifurcations, vessel segments, or compartments; edges encode flow, mechanical, or regulatory relationships. Node features typically include pressures, flows, cross-sectional areas, or patient covariates; edge features may include length, diameter, stiffness, or stenosis severity. Spatio-temporal signals arise when these quantities evolve over time, which is central to haemodynamics in digital twins~\cite{barbiero2021graph}.

\begin{figure}[H]
\centering
\includegraphics[width=\linewidth]{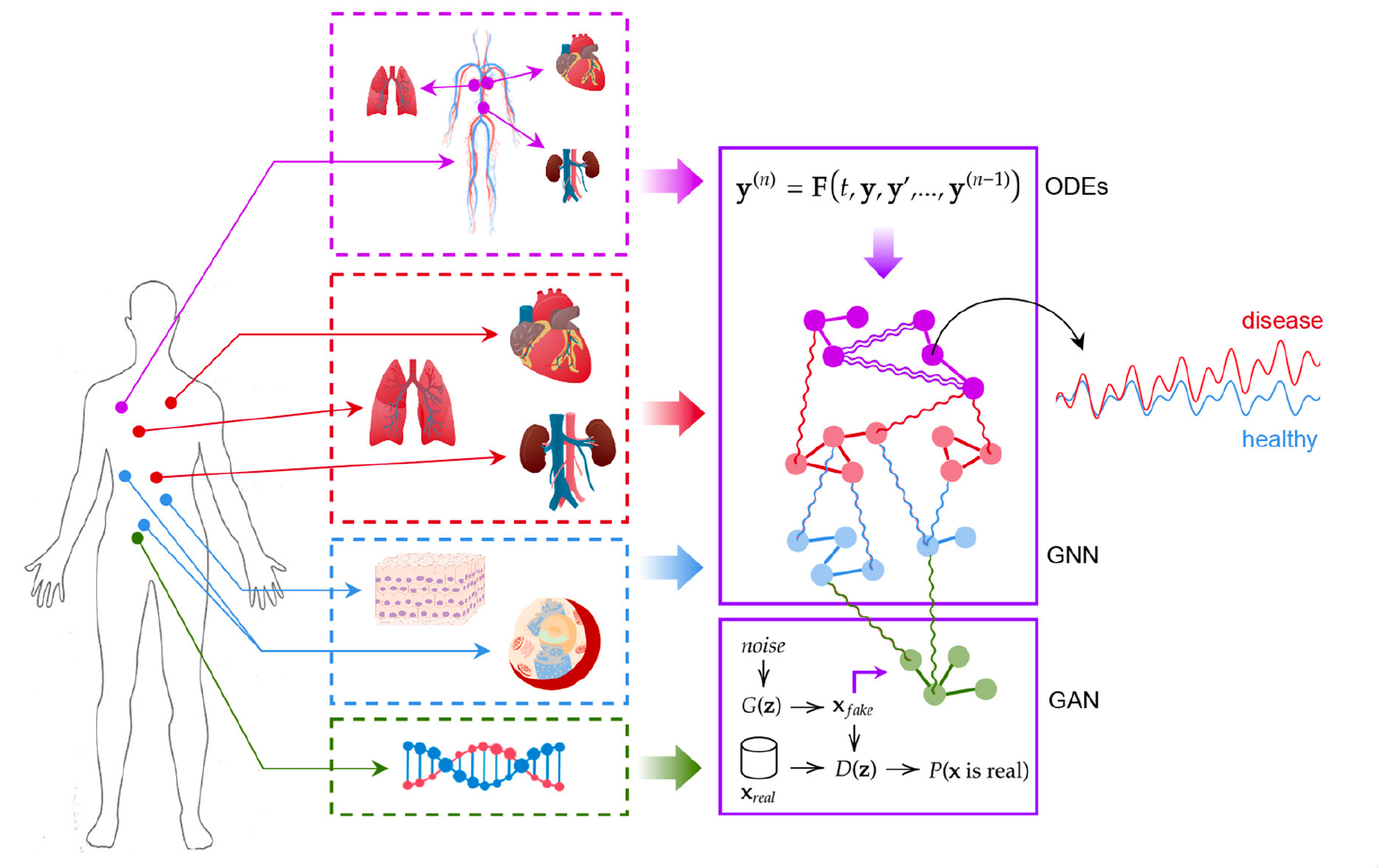}
\caption{Multi-scale cardiovascular digital twin integrating organ-, tissue-,
and cellular-level data using graph representations. Reproduced with permission
from Barbiero et al.~\cite{barbiero2021graph}.}
\label{fig:gnn_dt}
\end{figure}

Figure~\ref{fig:gnn_dt} illustrates graph neural networks within a broader cardiovascular digital twin architecture. In this framework, multi-scale physiological data, from organ-level systems such as the heart, lungs, and
kidneys, down to tissue- and cellular-level information, are mapped onto graph representations. Nodes may correspond to anatomical regions, vessel segments,
or functional compartments, while edges encode physiological coupling mechanisms such as flow continuity, pressure gradients, mechanical interaction, or regulatory feedback. By operating directly on these structured representations, GNNs enable scalable, topology-aware learning and provide a foundation for real-time, patient-specific cardiovascular digital twins that can integrate heterogeneous data sources and evolve with the patient state.

\subsection{GNN Architectures and Variants}
\label{sec:gnn_arch}

Most graph neural networks (GNNs) operate within the message-passing paradigm,
wherein each node aggregates information from its neighbours and subsequently
updates its internal state. Formally, the update at layer $k$ is:
\begin{equation}
\mathbf{h}_{v}^{(k)} = \mathrm{UPDATE}\!\left(
  \mathbf{h}_{v}^{(k-1)},\;
  \mathrm{AGGREGATE}\!\bigl(\{\mathbf{h}_{u}^{(k-1)} : u \in \mathcal{N}(v)\}\bigr)
\right),
\label{eq:mpnn}
\end{equation}
where $\mathbf{h}_{v}^{(k)}$ is the feature vector of node $v$ at layer $k$,
$\mathcal{N}(v)$ denotes its neighbourhood, and AGGREGATE and UPDATE are
learnable functions~\cite{paul2024systematic}.

Notable GNN variants relevant to cardiovascular applications include:
\begin{itemize}
  \item \emph{Graph Attention Networks (GATs)}, which assign learnable weights
    to neighbour contributions to preferentially highlight clinically relevant
    interactions, such as high-resistance vessel junctions in coronary tree
    modelling~\cite{shukla2022scalable}.
  \item \emph{GraphSAGE}, which enables inductive learning across graphs of
    variable size by sampling and aggregating neighbourhood features, useful
    when training data contain vascular networks of different anatomical
    complexity or when generalising to unseen patient
    geometries~\cite{paul2024systematic}.
  \item \emph{Message Passing Neural Networks (MPNNs)}, which provide a
    general unifying framework and have been extended to incorporate edge-level
    physics constraints such as flow conservation at
    bifurcations~\cite{paul2024systematic}.
  \item \emph{Spatio-temporal GNNs}, which integrate graph-based propagation
    with recurrent or convolutional temporal modules to capture the evolution
    of haemodynamic pressure and flow waveforms over time~\cite{iacovelli2024novel}.
  \item \emph{Neural ODE--GNN hybrids}, which couple continuous-time dynamics
    with graph-structured state representations, enabling simulation of
    haemodynamic waveforms as continuous trajectories rather than discretised
    sequences, an approach well-suited to the continuous temporal evolution of
    cardiovascular signals.
\end{itemize}

Physics-informed (or physics-aware) GNNs incorporate domain-specific prior knowledge by embedding constraints such as the conservation of mass at vascular junctions, monotonicity of resistive parameters, or structured parameterisations derived from reduced-order haemodynamic models. The inclusion of these inductive biases enhances numerical stability and promotes physiological plausibility, particularly in sparse-data or low-observability settings.

\subsection{Applications in Cardiovascular Digital Twin Frameworks}
\label{sec:gnn_applications}

Graph neural networks (GNNs) have been applied across a range of cardiovascular digital twin objectives: predicting pressure and flow waveforms across arterial networks, inferring unmeasured states from sparse sensor measurements, accelerating repeated simulations for uncertainty quantification or parameter inference, and modelling disease progression by learning how vessel properties evolve and affect network function. Additionally, GNNs serve as hybrid modules, complementing mechanistic cores (0D/1D models) with learnt corrections for boundary conditions or microcirculatory detail~\cite{barbiero2021graph,iacovelli2024novel}.

Concrete cardiovascular applications include:
\begin{itemize}
  \item \emph{Coronary tree pressure inference}: GNNs infer pressure and flow
    distributions across coronary arterial trees from sparse catheter
    measurements, where graph topology naturally encodes branching connectivity
    and enables prediction at unobserved vessel segments through message
    passing.
  \item \emph{Pulmonary arterial remodelling}: In pulmonary arterial
    hypertension, altered network-level resistance distributions drive disease
    progression in a fundamentally relational manner, making GNNs a natural
    modelling choice for capturing propagation of haemodynamic stress across
    the pulmonary tree.
  \item \emph{Stenosis propagation modelling}: Predicting how a localised
    luminal narrowing affects downstream pressure and flow throughout a network
    is inherently a graph-structured problem; GNN surrogates can evaluate these
    effects in milliseconds compared with minutes or hours for full CFD.
  \item \emph{Intervention planning}: GNN surrogates have been proposed to
    rapidly evaluate haemodynamic consequences of stent placement or bypass
    grafting across patient-specific vascular graphs, enabling virtual
    comparison of procedural strategies at low computational cost~\cite{wang20213d}.
\end{itemize}

GNNs complement both classical ML and PINNs. Classical ML scales well with
tabular data but often ignores network topology unless manually engineered.
GNNs, by contrast, exploit relational structure, generalise across variable
graph sizes, and efficiently perform local message aggregation~\cite{paul2024systematic}.
PINNs encode explicit differential equations, ensuring strong physics
consistency, but require known PDEs and may not scale naturally to large
networked systems. GNNs offer an intermediate solution: they can integrate
topology and approximate physics through physics-aware architectures, providing
scalable and interpretable models when combined with hybrid or surrogate
mechanistic frameworks~\cite{oloulade2021graph,you2020design}.

\subsection{GNNs for the Inverse Problem and Data Assimilation}
\label{sec:gnn_inverse}

Beyond forward simulation, GNNs are increasingly investigated as tools for the inverse problem central to digital twin personalisation. In this setting, a GNN may be trained to infer unmeasured node-level quantities, such as local flow resistance, vessel compliance, or boundary pressures, from partial observations at a subset of nodes. This capacity for graph-based state estimation is conceptually analogous to data assimilation, but operates through learned message passing rather than sequential Bayesian updating.

Compared with PINNs, which enforce governing PDEs explicitly at collocation points, GNNs offer greater scalability to large and irregular networks, at the cost of weaker guarantees of physical consistency unless physics-aware loss terms are incorporated. For model calibration in cardiovascular digital twins, GNNs and PINNs therefore represent complementary tools: PINNs may be preferred when the governing equations are well characterised and the domain is geometrically manageable, while GNNs are better suited to large networked systems where topology is a dominant inductive bias and relational data are sparse~\cite{barbiero2021graph,paul2024systematic}.

\subsection{Limitations and Open Challenges}
\label{sec:gnn_limits}

Designing cardiovascular GNNs involves several practical considerations. Node and edge feature selection should capture relevant physiological and anatomical descriptors. Network depth and aggregation functions must balance local and global influence, while temporal modelling is needed to capture dynamic haemodynamic waveforms. Oversmoothing, the tendency of deep GNNs to produce indistinguishable node representations, limits effective depth and remains an active research challenge~\cite{paul2024systematic}.

A fundamental limitation of standard GNNs in cardiovascular modelling is that
message-passing architectures do not inherently enforce physical conservation laws, such as mass conservation at vascular junctions or energy dissipation constraints, unless these are explicitly incorporated as loss terms or hard constraints. Unlike PINNs, where PDE residuals are minimised at collocation points, GNNs learn from data and may produce predictions that violate conservation principles under distribution shift or in extrapolative regimes. This limitation is particularly consequential in closed-loop haemodynamic models where mass imbalance at bifurcations can accumulate across network scales~\cite{paul2024systematic,barbiero2021graph}.

In summary, graph neural networks (GNNs) offer a scalable, topology-aware, and
modular computational framework for the representation and analysis of
cardiovascular networks. When integrated with mechanistic and physics-informed
modelling components, these architectures facilitate the development of hybrid
digital twins capable of real-time physiological monitoring, predictive
forecasting, and patient-specific simulation.

%=================================================================
\section{Hybrid and Multi-Paradigm Digital Twin Frameworks}
\label{sec:hybrid}

Cardiovascular digital twins increasingly adopt hybrid architectures that combine mechanistic modelling with artificial intelligence (AI). Purely biophysical models provide interpretability and physical consistency but are computationally intensive and difficult to personalise at scale. Conversely, data-driven models offer speed and scalability but may lack robustness and extrapolation capability. Hybrid frameworks integrate these paradigms to retain physiological grounding while enabling efficient inference.

One common strategy embeds AI surrogates within mechanistic solvers. Neural networks may replace computationally expensive 3D CFD or electromechanical components, while lower-dimensional (0D/1D) circulation models preserve interpretability. Such AI-accelerated frameworks enable rapid pressure, flow prediction and near-real-time simulation \cite{mynard2015one,zhang2020personalized}.  

AI also enhances parameter estimation and data assimilation. Machine learning can map clinical observations to latent physiological parameters or initialise optimisation-based calibration, reducing personalisation time while maintaining physical structure \cite{regazzoni2021combining}. Physics-informed learning further tightens integration by embedding governing equations within neural training, supporting data-efficient and physically consistent inference \cite{arzani2021data}.  

Recent developments emphasise modular twins, where subsystems use distinct paradigms: electrophysiology may remain biophysical, vascular transport may use graph neural networks (GNNs), and control layers may incorporate reinforcement learning \cite{kissas2023towards}. Such modularity supports extensibility while allowing validation at the subsystem level.

\subsection{Multi-Fidelity, Multi-Scale, and Real-Time Workflows}

Cardiovascular physiology spans scales from ion channels to organ-level haemodynamics. Multi-fidelity workflows couple low-resolution models for global dynamics with high-resolution simulations in regions of interest. Low-fidelity models provide boundary conditions and rapid screening; high-fidelity models are deployed selectively where detailed resolution is clinically necessary \cite{frangi2002three}.  

Multi-scale integration links cellular electrophysiology, tissue mechanics, and organ-level contraction within unified electromechanical frameworks \cite{kerckhoffs2006computational,trayanova2011whole}. AI-driven model reduction and operator learning enable substantial acceleration of high-dimensional solvers, making multi-scale coupling feasible in near-real-time contexts \cite{rodero2023advancing}.  

True digital twins extend beyond conventional simulation paradigms toward systems capable of continuous adaptation. Real-time digital twins assimilate streaming data from imaging modalities and wearable sensors to dynamically update internal states and model parameters \cite{dritsas2023efficient}. Closed-loop digital twins further integrate predictive modelling with optimisation frameworks, thereby supporting personalised therapy planning and advanced control strategies. Despite their potential, significant challenges persist, including issues related to data latency, model and sensor drift, as well as the stringent requirements associated with regulatory approval.

%=================================================================
\section{Verification, Validation, and Uncertainty Quantification}
\label{sec:vv_uq}

Robust clinical translation requires rigorous verification, validation, and
uncertainty quantification (VV\&UQ). Verification assesses numerical correctness, ensuring equations are solved accurately. For classical solvers, this includes mesh convergence and time-step analysis~\cite{frangi2002three,colebank2024guidelines}. For AI surrogates and
PINNs, verification involves residual consistency, comparison with high-fidelity
references, and physics-based diagnostics, as low training loss alone does not
guarantee accuracy~\cite{karniadakis2021physics}.

Validation evaluates whether predictions are clinically meaningful. In haemodynamics, this may involve comparison with catheter or MRI measurements; in electrophysiology, with ECG or activation maps~\cite{sun2014computational}. Because digital twins are personalised, validation must occur at both cohort and individual levels. External and multi-centre validation are essential to mitigate dataset bias~\cite{henglin2017machine}.

A critical distinction concerns the nature of the validation dataset.
Retrospective parameter fitting, in which model parameters are optimised to
reproduce measurements from the same patient, demonstrates personalisation
capacity but does not constitute independent validation. Validation on synthetic
datasets generated by high-fidelity simulators provides a controlled
environment for assessing numerical correctness but may not capture the
complexity and noise of real clinical data. Ex vivo or phantom-based validation
offers controlled ground truth while capturing some measurement realism.
Prospective validation in living patient cohorts, wherein the twin generates
predictions that are subsequently compared against independent clinical
outcomes, represents the highest evidential standard and remains comparatively
rare in the cardiovascular digital twin literature. The term \emph{patient-specific} should therefore be interpreted cautiously: it describes personalisation of model inputs or parameters, but does not by itself indicate prospective predictive validity~\cite{sel2024building,colebank2024guidelines}.

Uncertainty quantification (UQ) systematically characterises variability and
lack of knowledge in model parameters, experimental data, and model structure.
A foundational distinction is between \emph{aleatory} uncertainty, irreducible
variability arising from biological stochasticity, inter-patient heterogeneity,
and measurement noise, and \emph{epistemic} uncertainty, which reflects incomplete knowledge of model parameters, structure, or boundary conditions and is in principle reducible with additional data. In cardiovascular digital twins, both types are present simultaneously; failure to distinguish them can lead to miscalibrated predictions and inappropriate clinical confidence.

Biological systems inherently exhibit stochasticity and heterogeneity; consequently, neglecting uncertainty can yield overconfident and potentially misleading predictions~\cite{mirams2016uncertainty}. To mitigate this,
Bayesian calibration, ensemble-based modelling strategies, and probabilistic
inference frameworks are increasingly employed in both mechanistic and
AI-driven digital twins~\cite{regazzoni2021combining}.

Data assimilation frameworks constitute the mathematical backbone of true digital twin updating. Sequential methods such as the Ensemble Kalman Filter (EnKF) and Unscented Kalman Filter (UKF) propagate a state estimate and associated uncertainty forward in time using the model dynamics, then correct both when new observations become available. The EnKF, which approximates the covariance using an ensemble of model trajectories, has been applied to cardiovascular systems to jointly estimate haemodynamic states and model parameters from time-series pressure measurements. The UKF, using deterministic sigma points rather than random ensembles, offers computational advantages for lower-dimensional cardiovascular ODE models. Particle filters extend these approaches to non-Gaussian posteriors at substantially higher computational cost. Variational approaches, analogous to 4D-Var in meteorological data assimilation, formulate parameter estimation as a constrained optimisation problem over a time window, enabling batch updating from imaging or catheter data. In cardiovascular digital twin workflows, sequential Kalman-based updating is most practical for continuous physiological monitoring, while variational and Bayesian approaches are better suited to episodic clinical encounters~\cite{sun2014computational,regazzoni2021combining,duanmu2019one}.

Questions of both structural and practical identifiability, where distinct
parameter configurations can generate indistinguishable or highly similar model outputs, must be rigorously analysed and explicitly incorporated into all
stages of model formulation, calibration, and validation~\cite{mirams2016uncertainty,alonso2025biophysical}.

Benchmarking and reproducibility continue to represent critical and unresolved challenges. The availability of shared datasets, openly accessible and
well-documented implementations, as well as rigorous and transparent reporting practices constitute essential prerequisites for both regulatory acceptance and
sustained scientific advancement~\cite{sel2024building,paul2024systematic,colebank2024guidelines}.

%=================================================================
\section{Clinical and Translational Applications}
\label{sec:applications}

\subsection{Diagnosis and Risk Stratification}

Digital twins can facilitate personalised diagnosis by integrating anatomical,
physiological, and functional data into coherent, mechanistic predictive
models. Haemodynamic twins enable the inference of latent haemodynamic
variables, such as wall shear stress and pressure gradients, from
imaging-derived vascular geometries and sparse in vivo measurements~\cite{arzani2022machine}. Electrophysiological twins replicate
arrhythmogenic substrates and cardiac activation pathways, thereby supporting
mechanistic risk stratification that extends beyond conventional statistical
risk scores~\cite{trayanova2011whole}.

For instance, patient-specific electrophysiological digital twins have been
used to guide ablation target selection in patients with ventricular
tachycardia, with arrhythmia inducibility predicted pre-procedurally from
personalised computational models~\cite{trayanova2011whole}. In the vascular
domain, haemodynamic twins based on coronary computed tomography angiography
have enabled non-invasive estimation of fractional flow reserve, facilitating
lesion-specific treatment decisions without catheterisation~\cite{arzani2022machine}.

When coupled with machine learning or graph-based computational frameworks,
digital twins can enhance both scalability and interpretability relative to
purely statistical prediction models~\cite{sel2024building}. Their primary role
is to augment clinical decision-making by providing quantitative, patient-specific mechanistic insight.

\subsection{Intervention Planning and Personalised Medicine}

Digital twins facilitate in silico evaluation of therapeutic interventions,
including vascular stenting, valve replacement, and ablation procedures.
Through the simulation of haemodynamic and electrophysiological responses,
clinicians can systematically compare alternative treatment strategies prior to
performing the actual intervention~\cite{wang20213d}. In transcatheter aortic
valve replacement planning, for example, patient-specific haemodynamic
simulations are used to evaluate prosthesis sizing and deployment angles,
reducing the risk of paravalvular leak and conduction disturbance. The
incorporation of hybrid surrogate models further expedites the exploration of
multiple scenarios, which is particularly critical in time-sensitive clinical
decision-making contexts~\cite{kuang2024med}.

Longitudinal digital twins extend these capabilities toward predictive and
proactive healthcare. By assimilating repeated multimodal measurements over
time, such twins can iteratively update individual disease trajectories and
generate forecasts of treatment response~\cite{sel2024building}. At the
population level, personalised digital twins hold promise for enabling
precision medicine and optimising healthcare resource allocation; however,
rigorous approaches to bias identification, quantification, and mitigation are
essential to ensure equitable and reliable deployment.
%=================================================================
\section{Ethical, Regulatory, and Deployment Considerations}
\label{sec:regulatory}

Clinical implementation is contingent upon robust trust, transparency, and governance frameworks. While mechanistic model components inherently provide a degree of interpretability, AI-based modules necessitate dedicated
explainability methodologies and systematic uncertainty quantification to
support credible clinical decision-making and foster clinician confidence~\cite{kissas2023towards}. Furthermore, algorithmic bias and dataset
shift introduce substantial ethical concerns, particularly when training
datasets inadequately represent specific demographic or clinical subgroups,
thereby exacerbating disparities in model performance and health
outcomes~\cite{coorey2022health}.

Data privacy and security constitute fundamental prerequisites for modern
healthcare analytics. Federated and privacy-preserving machine learning
paradigms enable collaborative, multi-institutional model development while
ensuring that sensitive patient-level data remain protected and, where possible,
confined to local data custodians~\cite{rieke2020future}.

Regulators are increasingly classifying advanced digital twins as
software-as-a-medical-device (SaMD). Relevant regulatory frameworks include
the FDA's Digital Health Center of Excellence guidance on AI/ML-based SaMD,
the EU Medical Device Regulation (MDR 2017/745) which governs software meeting
the definition of a medical device, and the ISO 14971 standard for risk
management of medical devices. The ASME V\&V~40 standard specifically addresses
verification and validation of computational models used in medical device
submissions and provides a credibility assessment framework directly applicable
to cardiovascular digital twins~\cite{colebank2024guidelines}. It is therefore
critical to demonstrate safety, efficacy, and reproducibility, underpinned by
rigorous VV\&UQ~\cite{mirams2016uncertainty}. Adaptive learning systems add
further regulatory challenges, encouraging modular designs in which certified
mechanistic cores are integrated with AI components whose behaviour is strictly
constrained.

%=================================================================
\section{Challenges and Future Directions}
\label{sec:future}

Key methodological challenges encompass multi-scale integration, the treatment
of sparse and heterogeneous datasets, computational tractability, and the
achievement of robust generalisation across diverse conditions. It is
anticipated that hybrid modelling frameworks, integrating mechanistic,
physics-informed, and graph-based approaches, will play a predominant role in
subsequent methodological advances.

Emerging research trajectories encompass the development of foundation-style
pre-trained physiological models, the construction of scalable GNN
representations for complex vascular architectures, advances in the numerical
stability of PINNs for stiff multi-physics systems, and the design of adaptive,
closed-loop digital twins for predictive monitoring and control~\cite{sel2024building}.

An emerging trajectory of particular relevance is the development of foundation
models for biomedical and physiological data. Large pre-trained models,
analogous to those transforming natural language processing and general-purpose
vision, are beginning to be explored for multimodal clinical data integration,
ECG interpretation, and cardiac imaging analysis. Within the digital twin
paradigm, such models could serve as flexible priors for patient-specific
adaptation, reducing the data requirements for personalisation and enabling
rapid fine-tuning from small patient-specific datasets. However, their
integration with mechanistic cardiovascular models raises unresolved questions
about physical consistency, interpretability, and regulatory acceptance that
will require dedicated methodological investigation.

A pragmatic implementation roadmap prioritises the development of task-specific
digital twins, the application of rigorous VV\&UQ procedures, seamless
integration into existing computational and clinical workflows, and sustained
interdisciplinary collaboration. Furthermore, multi-centre validation and
methodological standardisation will be essential to achieve regulatory approval
and to ensure equitable, reproducible, and transparent deployment of these
technologies~\cite{sel2024building,colebank2024guidelines}.

%=================================================================
\section{Conclusions}
\label{sec:conclusions}

Cardiovascular digital twins integrate mechanistic modelling, machine learning, physics-informed learning, and graph-based representations to enable personalised and predictive simulations of cardiovascular function. Biophysical models provide interpretability and support causal inference, but they exhibit limitations in scalability and computational tractability. Data-driven approaches improve computational speed, adaptability, and scalability; however, they necessitate stringent safeguards to ensure robustness, generalisability, and fairness.  

Hybrid and multi-paradigm architectures constitute a unifying framework that combines physical consistency with computational efficiency and adaptive learning capabilities. Nonetheless, such technical advances must be accompanied by rigorous procedures for verification, validation, uncertainty quantification, and ethical as well as regulatory governance.  

As methodological maturity increases and regulatory frameworks evolve, cardiovascular digital twins are progressing from conceptual research instruments to clinically deployable technologies. Their ultimate impact will depend on their responsible integration into clinical workflows, transparent and standardised evaluation, and sustained interdisciplinary collaboration among engineering, clinical medicine, ethics, and regulatory science.

\section*{Author contributions}

%E.L. conceptualised the study, wrote, reviewed, and edited the original draft of the manuscript. 

Conceptualisation, E.L.; Writing-Original draft of the manuscript, E.L.; Writing-Reviewing \& Editing, F.C. The authors approved the final version of the manuscript.

\section*{Funding}

This research received no external funding. Grant number: Not applicable.

\section*{Competing interests}

The authors declare no competing interests.

%% Bibliography inlined below (pre-compiled from sn-bibliography.bib with
%% sn-mathphys-num.bst) so that no BibTeX run is required at submission.
%% Original command retained for reference:
%% \bibliography{sn-bibliography}

\begin{thebibliography}{80}
\ifx \bisbn   \undefined \def \bisbn  #1{ISBN #1}\fi
\ifx \binits  \undefined \def \binits#1{#1}\fi
\ifx \bauthor  \undefined \def \bauthor#1{#1}\fi
\ifx \batitle  \undefined \def \batitle#1{#1}\fi
\ifx \bjtitle  \undefined \def \bjtitle#1{#1}\fi
\ifx \bvolume  \undefined \def \bvolume#1{\textbf{#1}}\fi
\ifx \byear  \undefined \def \byear#1{#1}\fi
\ifx \bissue  \undefined \def \bissue#1{#1}\fi
\ifx \bfpage  \undefined \def \bfpage#1{#1}\fi
\ifx \blpage  \undefined \def \blpage #1{#1}\fi
\ifx \burl  \undefined \def \burl#1{\textsf{#1}}\fi
\ifx \doiurl  \undefined \def \doiurl#1{\url{https://doi.org/#1}}\fi
\ifx \betal  \undefined \def \betal{\textit{et al.}}\fi
\ifx \binstitute  \undefined \def \binstitute#1{#1}\fi
\ifx \binstitutionaled  \undefined \def \binstitutionaled#1{#1}\fi
\ifx \bctitle  \undefined \def \bctitle#1{#1}\fi
\ifx \beditor  \undefined \def \beditor#1{#1}\fi
\ifx \bpublisher  \undefined \def \bpublisher#1{#1}\fi
\ifx \bbtitle  \undefined \def \bbtitle#1{#1}\fi
\ifx \bedition  \undefined \def \bedition#1{#1}\fi
\ifx \bseriesno  \undefined \def \bseriesno#1{#1}\fi
\ifx \blocation  \undefined \def \blocation#1{#1}\fi
\ifx \bsertitle  \undefined \def \bsertitle#1{#1}\fi
\ifx \bsnm \undefined \def \bsnm#1{#1}\fi
\ifx \bsuffix \undefined \def \bsuffix#1{#1}\fi
\ifx \bparticle \undefined \def \bparticle#1{#1}\fi
\ifx \barticle \undefined \def \barticle#1{#1}\fi
\bibcommenthead
\ifx \bconfdate \undefined \def \bconfdate #1{#1}\fi
\ifx \botherref \undefined \def \botherref #1{#1}\fi
\ifx \url \undefined \def \url#1{\textsf{#1}}\fi
\ifx \bchapter \undefined \def \bchapter#1{#1}\fi
\ifx \bbook \undefined \def \bbook#1{#1}\fi
\ifx \bcomment \undefined \def \bcomment#1{#1}\fi
\ifx \oauthor \undefined \def \oauthor#1{#1}\fi
\ifx \citeauthoryear \undefined \def \citeauthoryear#1{#1}\fi
\ifx \endbibitem  \undefined \def \endbibitem {}\fi
\ifx \bconflocation  \undefined \def \bconflocation#1{#1}\fi
\ifx \arxivurl  \undefined \def \arxivurl#1{\textsf{#1}}\fi
\csname PreBibitemsHook\endcsname

%%% 1
\bibitem[\protect\citeauthoryear{Jeske}{2020}]{jeske2020digital}
\begin{botherref}
\oauthor{\bsnm{Jeske}, \binits{S.J.}}:
Digital twins in healthcare.
Paris Lodron University of Salzburg and Aalborg University in Copenhagen
(2020)
\end{botherref}
\endbibitem

%%% 2
\bibitem[\protect\citeauthoryear{Sel et~al.}{2024}]{sel2024building}
\begin{barticle}
\bauthor{\bsnm{Sel}, \binits{K.}},
\bauthor{\bsnm{Osman}, \binits{D.}},
\bauthor{\bsnm{Zare}, \binits{F.}},
\bauthor{\bsnm{Masoumi~Shahrbabak}, \binits{S.}},
\bauthor{\bsnm{Brattain}, \binits{L.}},
\bauthor{\bsnm{Hahn}, \binits{J.-O.}},
\bauthor{\bsnm{Inan}, \binits{O.T.}},
\bauthor{\bsnm{Mukkamala}, \binits{R.}},
\bauthor{\bsnm{Palmer}, \binits{J.}},
\bauthor{\bsnm{Paydarfar}, \binits{D.}}, \betal:
\batitle{Building digital twins for cardiovascular health: From principles to
  clinical impact}.
\bjtitle{Journal of the American Heart Association}
\bvolume{13}(\bissue{19}),
\bfpage{031981}
(\byear{2024})
\end{barticle}
\endbibitem

%%% 3
\bibitem[\protect\citeauthoryear{Kissas}{2023}]{kissas2023towards}
\begin{botherref}
\oauthor{\bsnm{Kissas}, \binits{G.}}:
Towards digital twins for cardiovascular flows: A hybrid machine learning and
  computational fluid dynamics approach.
PhD thesis,
University of Pennsylvania
(2023)
\end{botherref}
\endbibitem

%%% 4
\bibitem[\protect\citeauthoryear{Huang et~al.}{2024}]{huang2024application}
\begin{barticle}
\bauthor{\bsnm{Huang}, \binits{L.}},
\bauthor{\bsnm{Pan}, \binits{L.}},
\bauthor{\bsnm{Wu}, \binits{C.}},
\bauthor{\bsnm{Tian}, \binits{M.}},
\bauthor{\bsnm{Li}, \binits{Q.}},
\bauthor{\bsnm{Peng}, \binits{Y.}},
\bauthor{\bsnm{Li}, \binits{Q.}},
\bauthor{\bsnm{Li}, \binits{Y.}}:
\batitle{Application and development prospect of digital twin in the forensic
  identification of cardiovascular diseases}.
\bjtitle{Digital Medicine}
\bvolume{10}(\bissue{4}),
\bfpage{00013}
(\byear{2024})
\end{barticle}
\endbibitem

%%% 5
\bibitem[\protect\citeauthoryear{Canino et~al.}{2025}]{canino2025artificial}
\begin{barticle}
\bauthor{\bsnm{Canino}, \binits{G.}},
\bauthor{\bsnm{Di~Costanzo}, \binits{A.}},
\bauthor{\bsnm{Salerno}, \binits{N.}},
\bauthor{\bsnm{Leo}, \binits{I.}},
\bauthor{\bsnm{Cannataro}, \binits{M.}},
\bauthor{\bsnm{Guzzi}, \binits{P.H.}},
\bauthor{\bsnm{Veltri}, \binits{P.}},
\bauthor{\bsnm{Sorrentino}, \binits{S.}},
\bauthor{\bsnm{De~Rosa}, \binits{S.}},
\bauthor{\bsnm{Torella}, \binits{D.}}:
\batitle{Artificial intelligence in cardiac electrophysiology: a clinically
  oriented review with engineering primers}.
\bjtitle{Bioengineering}
\bvolume{12}(\bissue{10}),
\bfpage{1102}
(\byear{2025})
\end{barticle}
\endbibitem

%%% 6
\bibitem[\protect\citeauthoryear{Meijer et~al.}{2023}]{meijer2023digital}
\begin{barticle}
\bauthor{\bsnm{Meijer}, \binits{C.}},
\bauthor{\bsnm{Uh}, \binits{H.-W.}},
\bauthor{\bsnm{El~Bouhaddani}, \binits{S.}}:
\batitle{Digital twins in healthcare: Methodological challenges and
  opportunities}.
\bjtitle{Journal of personalized medicine}
\bvolume{13}(\bissue{10}),
\bfpage{1522}
(\byear{2023})
\end{barticle}
\endbibitem

%%% 7
\bibitem[\protect\citeauthoryear{Coorey et~al.}{2022}]{coorey2022health}
\begin{barticle}
\bauthor{\bsnm{Coorey}, \binits{G.}},
\bauthor{\bsnm{Figtree}, \binits{G.A.}},
\bauthor{\bsnm{Fletcher}, \binits{D.F.}},
\bauthor{\bsnm{Snelson}, \binits{V.J.}},
\bauthor{\bsnm{Vernon}, \binits{S.T.}},
\bauthor{\bsnm{Winlaw}, \binits{D.}},
\bauthor{\bsnm{Grieve}, \binits{S.M.}},
\bauthor{\bsnm{McEwan}, \binits{A.}},
\bauthor{\bsnm{Yang}, \binits{J.Y.H.}},
\bauthor{\bsnm{Qian}, \binits{P.}}, \betal:
\batitle{The health digital twin to tackle cardiovascular disease—a review of
  an emerging interdisciplinary field}.
\bjtitle{NPJ digital medicine}
\bvolume{5}(\bissue{1}),
\bfpage{126}
(\byear{2022})
\end{barticle}
\endbibitem

%%% 8
\bibitem[\protect\citeauthoryear{Strocchi et~al.}{2025}]{strocchi2025cardiac}
\begin{barticle}
\bauthor{\bsnm{Strocchi}, \binits{M.}},
\bauthor{\bsnm{Hammersley}, \binits{D.J.}},
\bauthor{\bsnm{Halliday}, \binits{B.P.}},
\bauthor{\bsnm{Prasad}, \binits{S.K.}},
\bauthor{\bsnm{Niederer}, \binits{S.A.}}:
\batitle{Cardiac digital twins: a tool to investigate the function and
  treatment of the diabetic heart}.
\bjtitle{Cardiovascular Diabetology}
\bvolume{24}(\bissue{1}),
\bfpage{293}
(\byear{2025})
\end{barticle}
\endbibitem

%%% 9
\bibitem[\protect\citeauthoryear{Zhao et~al.}{2025}]{zhao2025physics}
\begin{botherref}
\oauthor{\bsnm{Zhao}, \binits{A.}},
\oauthor{\bsnm{Fattahi}, \binits{D.}},
\oauthor{\bsnm{Hu}, \binits{X.}}:
Physics-informed neural networks for physiological signals processing and
  modeling: a narrative review.
Physiological measurement
(2025)
\end{botherref}
\endbibitem

%%% 10
\bibitem[\protect\citeauthoryear{Zhang et~al.}{2024}]{zhang2024concepts}
\begin{botherref}
\oauthor{\bsnm{Zhang}, \binits{K.}},
\oauthor{\bsnm{Zhou}, \binits{H.-Y.}},
\oauthor{\bsnm{Baptista-Hon}, \binits{D.T.}},
\oauthor{\bsnm{Gao}, \binits{Y.}},
\oauthor{\bsnm{Liu}, \binits{X.}},
\oauthor{\bsnm{Oermann}, \binits{E.}},
\oauthor{\bsnm{Xu}, \binits{S.}},
\oauthor{\bsnm{Jin}, \binits{S.}},
\oauthor{\bsnm{Zhang}, \binits{J.}},
\oauthor{\bsnm{Sun}, \binits{Z.}}, et al.:
Concepts and applications of digital twins in healthcare and medicine.
Patterns
\textbf{5}(8)
(2024)
\end{botherref}
\endbibitem

%%% 11
\bibitem[\protect\citeauthoryear{Arzani et~al.}{2022}]{arzani2022machine}
\begin{barticle}
\bauthor{\bsnm{Arzani}, \binits{A.}},
\bauthor{\bsnm{Wang}, \binits{J.-X.}},
\bauthor{\bsnm{Sacks}, \binits{M.S.}},
\bauthor{\bsnm{Shadden}, \binits{S.C.}}:
\batitle{Machine learning for cardiovascular biomechanics modeling: challenges
  and beyond}.
\bjtitle{Annals of Biomedical Engineering}
\bvolume{50}(\bissue{6}),
\bfpage{615}--\blpage{627}
(\byear{2022})
\end{barticle}
\endbibitem

%%% 12
\bibitem[\protect\citeauthoryear{Viola et~al.}{2023}]{viola2023gpu}
\begin{barticle}
\bauthor{\bsnm{Viola}, \binits{F.}},
\bauthor{\bsnm{Del~Corso}, \binits{G.}},
\bauthor{\bsnm{De~Paulis}, \binits{R.}},
\bauthor{\bsnm{Verzicco}, \binits{R.}}:
\batitle{Gpu accelerated digital twins of the human heart open new routes for
  cardiovascular research}.
\bjtitle{Scientific reports}
\bvolume{13}(\bissue{1}),
\bfpage{8230}
(\byear{2023})
\end{barticle}
\endbibitem

%%% 13
\bibitem[\protect\citeauthoryear{Li et~al.}{2024}]{li2024solving}
\begin{botherref}
\oauthor{\bsnm{Li}, \binits{L.}},
\oauthor{\bsnm{Camps}, \binits{J.}},
\oauthor{\bsnm{Rodriguez}, \binits{B.}},
\oauthor{\bsnm{Grau}, \binits{V.}}:
Solving the inverse problem of electrocardiography for cardiac digital twins: A
  survey.
IEEE Reviews in Biomedical Engineering
(2024)
\end{botherref}
\endbibitem

%%% 14
\bibitem[\protect\citeauthoryear{Kerckhoffs
  et~al.}{2006}]{kerckhoffs2006computational}
\begin{barticle}
\bauthor{\bsnm{Kerckhoffs}, \binits{R.C.}},
\bauthor{\bsnm{Healy}, \binits{S.N.}},
\bauthor{\bsnm{Usyk}, \binits{T.P.}},
\bauthor{\bsnm{McCULLOCH}, \binits{A.D.}}:
\batitle{Computational methods for cardiac electromechanics}.
\bjtitle{Proceedings of the IEEE}
\bvolume{94}(\bissue{4}),
\bfpage{769}--\blpage{783}
(\byear{2006})
\end{barticle}
\endbibitem

%%% 15
\bibitem[\protect\citeauthoryear{Xu and Wang}{2025}]{xu2025cardiac}
\begin{barticle}
\bauthor{\bsnm{Xu}, \binits{J.}},
\bauthor{\bsnm{Wang}, \binits{F.}}:
\batitle{Cardiac mechano-electrical-fluid interaction: a brief review of recent
  advances}.
\bjtitle{Eng}
\bvolume{6}(\bissue{8}),
\bfpage{168}
(\byear{2025})
\end{barticle}
\endbibitem

%%% 16
\bibitem[\protect\citeauthoryear{Tes{\'a}n
  et~al.}{2025}]{tesan2025thermodynamics}
\begin{botherref}
\oauthor{\bsnm{Tes{\'a}n}, \binits{L.}},
\oauthor{\bsnm{Gonz{\'a}lez}, \binits{D.}},
\oauthor{\bsnm{Martins}, \binits{P.}},
\oauthor{\bsnm{Cueto}, \binits{E.}}:
Thermodynamics-informed graph neural networks for real-time simulation of
  digital human twins.
Computational Mechanics,
1--22
(2025)
\end{botherref}
\endbibitem

%%% 17
\bibitem[\protect\citeauthoryear{Bewig}{2025}]{bewig2025cardiovascular}
\begin{botherref}
\oauthor{\bsnm{Bewig}, \binits{N.}}:
Cardiovascular Digital Twins from Time-Resolved CT
(2025)
\end{botherref}
\endbibitem

%%% 18
\bibitem[\protect\citeauthoryear{Zhang et~al.}{2020}]{zhang2020personalized}
\begin{barticle}
\bauthor{\bsnm{Zhang}, \binits{X.}},
\bauthor{\bsnm{Wu}, \binits{D.}},
\bauthor{\bsnm{Miao}, \binits{F.}},
\bauthor{\bsnm{Liu}, \binits{H.}},
\bauthor{\bsnm{Li}, \binits{Y.}}:
\batitle{Personalized hemodynamic modeling of the human cardiovascular system:
  a reduced-order computing model}.
\bjtitle{IEEE Transactions on Biomedical Engineering}
\bvolume{67}(\bissue{10}),
\bfpage{2754}--\blpage{2764}
(\byear{2020})
\end{barticle}
\endbibitem

%%% 19
\bibitem[\protect\citeauthoryear{Gray and Pathmanathan}{2018}]{gray2018patient}
\begin{barticle}
\bauthor{\bsnm{Gray}, \binits{R.A.}},
\bauthor{\bsnm{Pathmanathan}, \binits{P.}}:
\batitle{Patient-specific cardiovascular computational modeling: diversity of
  personalization and challenges}.
\bjtitle{Journal of cardiovascular translational research}
\bvolume{11}(\bissue{2}),
\bfpage{80}--\blpage{88}
(\byear{2018})
\end{barticle}
\endbibitem

%%% 20
\bibitem[\protect\citeauthoryear{Rudnicka et~al.}{2024}]{rudnicka2024cardiac}
\begin{barticle}
\bauthor{\bsnm{Rudnicka}, \binits{Z.}},
\bauthor{\bsnm{Proniewska}, \binits{K.}},
\bauthor{\bsnm{Perkins}, \binits{M.}},
\bauthor{\bsnm{Pregowska}, \binits{A.}}:
\batitle{Cardiac healthcare digital twins supported by artificial
  intelligence-based algorithms and extended reality—a systematic review}.
\bjtitle{Electronics}
\bvolume{13}(\bissue{5}),
\bfpage{866}
(\byear{2024})
\end{barticle}
\endbibitem

%%% 21
\bibitem[\protect\citeauthoryear{Mou et~al.}{2015}]{mou2015exploring}
\begin{barticle}
\bauthor{\bsnm{Mou}, \binits{Y.A.}},
\bauthor{\bsnm{Bollensdorff}, \binits{C.}},
\bauthor{\bsnm{Cazorla}, \binits{O.}},
\bauthor{\bsnm{Magdi}, \binits{Y.}},
\bauthor{\bsnm{De~Tombe}, \binits{P.P.}}:
\batitle{Exploring cardiac biophysical properties}.
\bjtitle{Global Cardiology Science and Practice}
\bvolume{2015}(\bissue{1}),
\bfpage{10}
(\byear{2015})
\end{barticle}
\endbibitem

%%% 22
\bibitem[\protect\citeauthoryear{Naik and
  Bhathawala}{2017}]{naik2017mathematical}
\begin{barticle}
\bauthor{\bsnm{Naik}, \binits{K.}},
\bauthor{\bsnm{Bhathawala}, \binits{P.}}:
\batitle{Mathematical modeling of human cardiovascular system: A lumped
  parameter approach and simulation}.
\bjtitle{International Journal of Mathematical, Computational, Physical,
  Electrical and Computer Engineering}
\bvolume{11}(\bissue{2}),
\bfpage{72}--\blpage{84}
(\byear{2017})
\end{barticle}
\endbibitem

%%% 23
\bibitem[\protect\citeauthoryear{Shi}{2013}]{shi2013lumped}
\begin{botherref}
\oauthor{\bsnm{Shi}, \binits{Y.}}:
Lumped-parameter modelling of cardiovascular system dynamics under different
  healthy and diseased conditions.
PhD thesis,
University of Sheffield
(2013)
\end{botherref}
\endbibitem

%%% 24
\bibitem[\protect\citeauthoryear{Mynard and Smolich}{2015}]{mynard2015one}
\begin{barticle}
\bauthor{\bsnm{Mynard}, \binits{J.P.}},
\bauthor{\bsnm{Smolich}, \binits{J.J.}}:
\batitle{One-dimensional haemodynamic modeling and wave dynamics in the entire
  adult circulation}.
\bjtitle{Annals of biomedical engineering}
\bvolume{43}(\bissue{6}),
\bfpage{1443}--\blpage{1460}
(\byear{2015})
\end{barticle}
\endbibitem

%%% 25
\bibitem[\protect\citeauthoryear{Mirams et~al.}{2016}]{mirams2016uncertainty}
\begin{barticle}
\bauthor{\bsnm{Mirams}, \binits{G.R.}},
\bauthor{\bsnm{Pathmanathan}, \binits{P.}},
\bauthor{\bsnm{Gray}, \binits{R.A.}},
\bauthor{\bsnm{Challenor}, \binits{P.}},
\bauthor{\bsnm{Clayton}, \binits{R.H.}}:
\batitle{Uncertainty and variability in computational and mathematical models
  of cardiac physiology}.
\bjtitle{The Journal of physiology}
\bvolume{594}(\bissue{23}),
\bfpage{6833}--\blpage{6847}
(\byear{2016})
\end{barticle}
\endbibitem

%%% 26
\bibitem[\protect\citeauthoryear{Trayanova}{2011}]{trayanova2011whole}
\begin{barticle}
\bauthor{\bsnm{Trayanova}, \binits{N.A.}}:
\batitle{Whole-heart modeling: applications to cardiac electrophysiology and
  electromechanics}.
\bjtitle{Circulation research}
\bvolume{108}(\bissue{1}),
\bfpage{113}--\blpage{128}
(\byear{2011})
\end{barticle}
\endbibitem

%%% 27
\bibitem[\protect\citeauthoryear{Shi et~al.}{2011}]{shi2011review}
\begin{barticle}
\bauthor{\bsnm{Shi}, \binits{Y.}},
\bauthor{\bsnm{Lawford}, \binits{P.}},
\bauthor{\bsnm{Hose}, \binits{R.}}:
\batitle{Review of zero-d and 1-d models of blood flow in the cardiovascular
  system}.
\bjtitle{Biomedical engineering online}
\bvolume{10}(\bissue{1}),
\bfpage{33}
(\byear{2011})
\end{barticle}
\endbibitem

%%% 28
\bibitem[\protect\citeauthoryear{Larrabide et~al.}{2012}]{larrabide2012hemolab}
\begin{barticle}
\bauthor{\bsnm{Larrabide}, \binits{I.}},
\bauthor{\bsnm{Blanco}, \binits{P.J.}},
\bauthor{\bsnm{Urquiza}, \binits{S.A.}},
\bauthor{\bsnm{Dari}, \binits{E.A.}},
\bauthor{\bsnm{V{\'e}nere}, \binits{M.J.}},
\bauthor{\bsnm{Silva}, \binits{N.d.S.}},
\bauthor{\bsnm{Feij{\'o}o}, \binits{R.A.}}:
\batitle{Hemolab--hemodynamics modelling laboratory: An application for
  modelling the human cardiovascular system}.
\bjtitle{Computers in biology and medicine}
\bvolume{42}(\bissue{10}),
\bfpage{993}--\blpage{1004}
(\byear{2012})
\end{barticle}
\endbibitem

%%% 29
\bibitem[\protect\citeauthoryear{Cai et~al.}{2024}]{cai2024lumped}
\begin{barticle}
\bauthor{\bsnm{Cai}, \binits{L.}},
\bauthor{\bsnm{Zhong}, \binits{Q.}},
\bauthor{\bsnm{Xu}, \binits{J.}},
\bauthor{\bsnm{Huang}, \binits{Y.}},
\bauthor{\bsnm{Gao}, \binits{H.}}:
\batitle{A lumped parameter model for evaluating coronary artery blood supply
  capacity}.
\bjtitle{Mathematical Biosciences and Engineering}
\bvolume{21}(\bissue{4}),
\bfpage{5838}--\blpage{5862}
(\byear{2024})
\end{barticle}
\endbibitem

%%% 30
\bibitem[\protect\citeauthoryear{Frangi et~al.}{2002}]{frangi2002three}
\begin{barticle}
\bauthor{\bsnm{Frangi}, \binits{A.F.}},
\bauthor{\bsnm{Niessen}, \binits{W.J.}},
\bauthor{\bsnm{Viergever}, \binits{M.A.}}:
\batitle{Three-dimensional modeling for functional analysis of cardiac images,
  a review}.
\bjtitle{IEEE transactions on medical imaging}
\bvolume{20}(\bissue{1}),
\bfpage{2}--\blpage{5}
(\byear{2002})
\end{barticle}
\endbibitem

%%% 31
\bibitem[\protect\citeauthoryear{Colebank
  et~al.}{2024}]{colebank2024guidelines}
\begin{barticle}
\bauthor{\bsnm{Colebank}, \binits{M.J.}},
\bauthor{\bsnm{Oomen}, \binits{P.A.}},
\bauthor{\bsnm{Witzenburg}, \binits{C.M.}},
\bauthor{\bsnm{Grosberg}, \binits{A.}},
\bauthor{\bsnm{Beard}, \binits{D.A.}},
\bauthor{\bsnm{Husmeier}, \binits{D.}},
\bauthor{\bsnm{Olufsen}, \binits{M.S.}},
\bauthor{\bsnm{Chesler}, \binits{N.C.}}:
\batitle{Guidelines for mechanistic modeling and analysis in cardiovascular
  research}.
\bjtitle{American Journal of Physiology-Heart and Circulatory Physiology}
\bvolume{327}(\bissue{2}),
\bfpage{473}--\blpage{503}
(\byear{2024})
\end{barticle}
\endbibitem

%%% 32
\bibitem[\protect\citeauthoryear{Smith et~al.}{2011}]{smith2011euheart}
\begin{barticle}
\bauthor{\bsnm{Smith}, \binits{N.}},
\bauthor{\bsnm{Vecchi}, \binits{A.}},
\bauthor{\bsnm{McCormick}, \binits{M.}},
\bauthor{\bsnm{Nordsletten}, \binits{D.}},
\bauthor{\bsnm{Camara}, \binits{O.}},
\bauthor{\bsnm{Frangi}, \binits{A.F.}},
\bauthor{\bsnm{Delingette}, \binits{H.}},
\bauthor{\bsnm{Sermesant}, \binits{M.}},
\bauthor{\bsnm{Relan}, \binits{J.}},
\bauthor{\bsnm{Ayache}, \binits{N.}}, \betal:
\batitle{euheart: personalized and integrated cardiac care using
  patient-specific cardiovascular modelling}.
\bjtitle{Interface focus}
\bvolume{1}(\bissue{3}),
\bfpage{349}--\blpage{364}
(\byear{2011})
\end{barticle}
\endbibitem

%%% 33
\bibitem[\protect\citeauthoryear{Chen et~al.}{2024}]{chen2024coupling}
\begin{barticle}
\bauthor{\bsnm{Chen}, \binits{R.}},
\bauthor{\bsnm{Cui}, \binits{J.}},
\bauthor{\bsnm{Li}, \binits{S.}},
\bauthor{\bsnm{Hao}, \binits{A.}}:
\batitle{A coupling physics model for real-time 4d simulation of cardiac
  electromechanics}.
\bjtitle{Computer-Aided Design}
\bvolume{175},
\bfpage{103747}
(\byear{2024})
\end{barticle}
\endbibitem

%%% 34
\bibitem[\protect\citeauthoryear{Sun et~al.}{2014}]{sun2014computational}
\begin{barticle}
\bauthor{\bsnm{Sun}, \binits{W.}},
\bauthor{\bsnm{Martin}, \binits{C.}},
\bauthor{\bsnm{Pham}, \binits{T.}}:
\batitle{Computational modeling of cardiac valve function and intervention}.
\bjtitle{Annual review of biomedical engineering}
\bvolume{16}(\bissue{1}),
\bfpage{53}--\blpage{76}
(\byear{2014})
\end{barticle}
\endbibitem

%%% 35
\bibitem[\protect\citeauthoryear{Bucelli
  et~al.}{2023}]{bucelli2023mathematical}
\begin{barticle}
\bauthor{\bsnm{Bucelli}, \binits{M.}},
\bauthor{\bsnm{Zingaro}, \binits{A.}},
\bauthor{\bsnm{Africa}, \binits{P.C.}},
\bauthor{\bsnm{Fumagalli}, \binits{I.}},
\bauthor{\bsnm{Dede'}, \binits{L.}},
\bauthor{\bsnm{Quarteroni}, \binits{A.}}:
\batitle{A mathematical model that integrates cardiac electrophysiology,
  mechanics, and fluid dynamics: Application to the human left heart}.
\bjtitle{International journal for numerical methods in biomedical engineering}
\bvolume{39}(\bissue{3}),
\bfpage{3678}
(\byear{2023})
\end{barticle}
\endbibitem

%%% 36
\bibitem[\protect\citeauthoryear{Gul}{2016}]{gul2016mathematical}
\begin{botherref}
\oauthor{\bsnm{Gul}, \binits{R.}}:
Mathematical modeling and sensitivity analysis of lumped-parameter model of the
  human cardiovascular system.
PhD thesis,
Freie Universit\"at Berlin,
Berlin, Germany
(2016).
\url{https://refubium.fu-berlin.de/handle/fub188/9472}
\end{botherref}
\endbibitem

%%% 37
\bibitem[\protect\citeauthoryear{Duanmu et~al.}{2019}]{duanmu2019one}
\begin{barticle}
\bauthor{\bsnm{Duanmu}, \binits{Z.}},
\bauthor{\bsnm{Chen}, \binits{W.}},
\bauthor{\bsnm{Gao}, \binits{H.}},
\bauthor{\bsnm{Yang}, \binits{X.}},
\bauthor{\bsnm{Luo}, \binits{X.}},
\bauthor{\bsnm{Hill}, \binits{N.A.}}:
\batitle{A one-dimensional hemodynamic model of the coronary arterial tree}.
\bjtitle{Frontiers in Physiology}
\bvolume{10},
\bfpage{853}
(\byear{2019})
\end{barticle}
\endbibitem

%%% 38
\bibitem[\protect\citeauthoryear{Rodero et~al.}{2023}]{rodero2023advancing}
\begin{barticle}
\bauthor{\bsnm{Rodero}, \binits{C.}},
\bauthor{\bsnm{Baptiste}, \binits{T.M.}},
\bauthor{\bsnm{Barrows}, \binits{R.K.}},
\bauthor{\bsnm{Lewalle}, \binits{A.}},
\bauthor{\bsnm{Niederer}, \binits{S.A.}},
\bauthor{\bsnm{Strocchi}, \binits{M.}}:
\batitle{Advancing clinical translation of cardiac biomechanics models: a
  comprehensive review, applications and future pathways}.
\bjtitle{Frontiers in physics}
\bvolume{11},
\bfpage{1306210}
(\byear{2023})
\end{barticle}
\endbibitem

%%% 39
\bibitem[\protect\citeauthoryear{Garber et~al.}{2022}]{garber2022critical}
\begin{barticle}
\bauthor{\bsnm{Garber}, \binits{L.}},
\bauthor{\bsnm{Khodaei}, \binits{S.}},
\bauthor{\bsnm{Keshavarz-Motamed}, \binits{Z.}}:
\batitle{The critical role of lumped parameter models in patient-specific
  cardiovascular simulations}.
\bjtitle{Archives of computational methods in engineering}
\bvolume{29}(\bissue{5}),
\bfpage{2977}--\blpage{3000}
(\byear{2022})
\end{barticle}
\endbibitem

%%% 40
\bibitem[\protect\citeauthoryear{Alonso et~al.}{2025}]{alonso2025biophysical}
\begin{barticle}
\bauthor{\bsnm{Alonso}, \binits{S.}},
\bauthor{\bsnm{Alvarez-Lacalle}, \binits{E.}},
\bauthor{\bsnm{Bragard}, \binits{J.}},
\bauthor{\bsnm{Echebarria}, \binits{B.}}:
\batitle{Biophysical modeling of cardiac cells: From ion channels to tissue}.
\bjtitle{Biophysica}
\bvolume{5}(\bissue{1}),
\bfpage{5}
(\byear{2025})
\end{barticle}
\endbibitem

%%% 41
\bibitem[\protect\citeauthoryear{Wang et~al.}{2021}]{wang20213d}
\begin{barticle}
\bauthor{\bsnm{Wang}, \binits{D.D.}},
\bauthor{\bsnm{Qian}, \binits{Z.}},
\bauthor{\bsnm{Vukicevic}, \binits{M.}},
\bauthor{\bsnm{Engelhardt}, \binits{S.}},
\bauthor{\bsnm{Kheradvar}, \binits{A.}},
\bauthor{\bsnm{Zhang}, \binits{C.}},
\bauthor{\bsnm{Little}, \binits{S.H.}},
\bauthor{\bsnm{Verjans}, \binits{J.}},
\bauthor{\bsnm{Comaniciu}, \binits{D.}},
\bauthor{\bsnm{O’Neill}, \binits{W.W.}}, \betal:
\batitle{3d printing, computational modeling, and artificial intelligence for
  structural heart disease}.
\bjtitle{Cardiovascular Imaging}
\bvolume{14}(\bissue{1}),
\bfpage{41}--\blpage{60}
(\byear{2021})
\end{barticle}
\endbibitem

%%% 42
\bibitem[\protect\citeauthoryear{Dinh et~al.}{2019}]{dinh2019data}
\begin{barticle}
\bauthor{\bsnm{Dinh}, \binits{A.}},
\bauthor{\bsnm{Miertschin}, \binits{S.}},
\bauthor{\bsnm{Young}, \binits{A.}},
\bauthor{\bsnm{Mohanty}, \binits{S.D.}}:
\batitle{A data-driven approach to predicting diabetes and cardiovascular
  disease with machine learning}.
\bjtitle{BMC medical informatics and decision making}
\bvolume{19}(\bissue{1}),
\bfpage{1}--\blpage{15}
(\byear{2019})
\end{barticle}
\endbibitem

%%% 43
\bibitem[\protect\citeauthoryear{van Osta et~al.}{2025}]{van2025individual}
\begin{botherref}
\oauthor{\bsnm{Osta}, \binits{N.}},
\oauthor{\bsnm{Loon}, \binits{T.}},
\oauthor{\bsnm{Lumens}, \binits{J.}}:
Individual hearts: computational models for improved management of
  cardiovascular disease.
Heart
(2025)
\end{botherref}
\endbibitem

%%% 44
\bibitem[\protect\citeauthoryear{Henglin et~al.}{2017}]{henglin2017machine}
\begin{barticle}
\bauthor{\bsnm{Henglin}, \binits{M.}},
\bauthor{\bsnm{Stein}, \binits{G.}},
\bauthor{\bsnm{Hushcha}, \binits{P.V.}},
\bauthor{\bsnm{Snoek}, \binits{J.}},
\bauthor{\bsnm{Wiltschko}, \binits{A.B.}},
\bauthor{\bsnm{Cheng}, \binits{S.}}:
\batitle{Machine learning approaches in cardiovascular imaging}.
\bjtitle{Circulation: Cardiovascular Imaging}
\bvolume{10}(\bissue{10}),
\bfpage{005614}
(\byear{2017})
\end{barticle}
\endbibitem

%%% 45
\bibitem[\protect\citeauthoryear{Arzani and Dawson}{2021}]{arzani2021data}
\begin{barticle}
\bauthor{\bsnm{Arzani}, \binits{A.}},
\bauthor{\bsnm{Dawson}, \binits{S.T.}}:
\batitle{Data-driven cardiovascular flow modelling: examples and
  opportunities}.
\bjtitle{Journal of the Royal Society Interface}
\bvolume{18}(\bissue{175}),
\bfpage{20200802}
(\byear{2021})
\end{barticle}
\endbibitem

%%% 46
\bibitem[\protect\citeauthoryear{Regazzoni
  et~al.}{2021}]{regazzoni2021combining}
\begin{barticle}
\bauthor{\bsnm{Regazzoni}, \binits{F.}},
\bauthor{\bsnm{Chapelle}, \binits{D.}},
\bauthor{\bsnm{Moireau}, \binits{P.}}:
\batitle{Combining data assimilation and machine learning to build data-driven
  models for unknown long time dynamics—applications in cardiovascular
  modeling}.
\bjtitle{International Journal for Numerical Methods in Biomedical Engineering}
\bvolume{37}(\bissue{7}),
\bfpage{3471}
(\byear{2021})
\end{barticle}
\endbibitem

%%% 47
\bibitem[\protect\citeauthoryear{Bauer et~al.}{2023}]{bauer2023data}
\begin{botherref}
\oauthor{\bsnm{Bauer}, \binits{R.}},
\oauthor{\bsnm{Cicero}, \binits{A.F.G.}},
\oauthor{\bsnm{Manca}, \binits{M.}}:
Data driven and model based computational futures in cardiovascular practice.
Frontiers Media SA
(2023)
\end{botherref}
\endbibitem

%%% 48
\bibitem[\protect\citeauthoryear{Gandin
  et~al.}{2021}]{gandin2021interpretability}
\begin{barticle}
\bauthor{\bsnm{Gandin}, \binits{I.}},
\bauthor{\bsnm{Scagnetto}, \binits{A.}},
\bauthor{\bsnm{Romani}, \binits{S.}},
\bauthor{\bsnm{Barbati}, \binits{G.}}:
\batitle{Interpretability of time-series deep learning models: A study in
  cardiovascular patients admitted to intensive care unit}.
\bjtitle{Journal of biomedical informatics}
\bvolume{121},
\bfpage{103876}
(\byear{2021})
\end{barticle}
\endbibitem

%%% 49
\bibitem[\protect\citeauthoryear{Dritsas and
  Trigka}{2023}]{dritsas2023efficient}
\begin{barticle}
\bauthor{\bsnm{Dritsas}, \binits{E.}},
\bauthor{\bsnm{Trigka}, \binits{M.}}:
\batitle{Efficient data-driven machine learning models for cardiovascular
  diseases risk prediction}.
\bjtitle{Sensors}
\bvolume{23}(\bissue{3}),
\bfpage{1161}
(\byear{2023})
\end{barticle}
\endbibitem

%%% 50
\bibitem[\protect\citeauthoryear{Kissi et~al.}{2025}]{kissi2025data}
\begin{barticle}
\bauthor{\bsnm{Kissi}, \binits{S.A.}},
\bauthor{\bsnm{Talukder}, \binits{M.G.M.}},
\bauthor{\bsnm{Iqbal}, \binits{M.Z.}}:
\batitle{Data-driven predictive modelling of lifestyle risk factors for
  cardiovascular health}.
\bjtitle{Electronics}
\bvolume{14}(\bissue{14}),
\bfpage{2906}
(\byear{2025})
\end{barticle}
\endbibitem

%%% 51
\bibitem[\protect\citeauthoryear{Barzegar~Gerdroodbary and
  Salavatidezfouli}{2025}]{barzegar2025predictive}
\begin{barticle}
\bauthor{\bsnm{Barzegar~Gerdroodbary}, \binits{M.}},
\bauthor{\bsnm{Salavatidezfouli}, \binits{S.}}:
\batitle{A predictive surrogate model of blood haemodynamics for
  patient-specific carotid artery stenosis}.
\bjtitle{Journal of the Royal Society Interface}
\bvolume{22}(\bissue{224}),
\bfpage{20240774}
(\byear{2025})
\end{barticle}
\endbibitem

%%% 52
\bibitem[\protect\citeauthoryear{Morid et~al.}{2023}]{morid2023time}
\begin{barticle}
\bauthor{\bsnm{Morid}, \binits{M.A.}},
\bauthor{\bsnm{Sheng}, \binits{O.R.L.}},
\bauthor{\bsnm{Dunbar}, \binits{J.}}:
\batitle{Time series prediction using deep learning methods in healthcare}.
\bjtitle{ACM Transactions on Management Information Systems}
\bvolume{14}(\bissue{1}),
\bfpage{1}--\blpage{29}
(\byear{2023})
\end{barticle}
\endbibitem

%%% 53
\bibitem[\protect\citeauthoryear{Shameer et~al.}{2018}]{shameer2018machine}
\begin{barticle}
\bauthor{\bsnm{Shameer}, \binits{K.}},
\bauthor{\bsnm{Johnson}, \binits{K.W.}},
\bauthor{\bsnm{Glicksberg}, \binits{B.S.}},
\bauthor{\bsnm{Dudley}, \binits{J.T.}},
\bauthor{\bsnm{Sengupta}, \binits{P.P.}}:
\batitle{Machine learning in cardiovascular medicine: are we there yet?}
\bjtitle{Heart}
\bvolume{104}(\bissue{14}),
\bfpage{1156}--\blpage{1164}
(\byear{2018})
\end{barticle}
\endbibitem

%%% 54
\bibitem[\protect\citeauthoryear{Prabhu et~al.}{2023}]{prabhu2023data}
\begin{barticle}
\bauthor{\bsnm{Prabhu}, \binits{S.}},
\bauthor{\bsnm{Rangarajan}, \binits{S.}},
\bauthor{\bsnm{Kothare}, \binits{M.}}:
\batitle{Data-driven discovery of sparse dynamical model of cardiovascular
  system for model predictive control}.
\bjtitle{Computers in biology and medicine}
\bvolume{166},
\bfpage{107513}
(\byear{2023})
\end{barticle}
\endbibitem

%%% 55
\bibitem[\protect\citeauthoryear{Gerdroodbary and
  Salavatidezfouli}{2025}]{gerdroodbary2025predictive}
\begin{barticle}
\bauthor{\bsnm{Gerdroodbary}, \binits{M.B.}},
\bauthor{\bsnm{Salavatidezfouli}, \binits{S.}}:
\batitle{A predictive surrogate model based on linear and nonlinear solution
  manifold reduction in cardiovascular fsi: A comparative study}.
\bjtitle{Computers in Biology and Medicine}
\bvolume{189},
\bfpage{109959}
(\byear{2025})
\end{barticle}
\endbibitem

%%% 56
\bibitem[\protect\citeauthoryear{Dozen et~al.}{2020}]{dozen2020image}
\begin{barticle}
\bauthor{\bsnm{Dozen}, \binits{A.}},
\bauthor{\bsnm{Komatsu}, \binits{M.}},
\bauthor{\bsnm{Sakai}, \binits{A.}},
\bauthor{\bsnm{Komatsu}, \binits{R.}},
\bauthor{\bsnm{Shozu}, \binits{K.}},
\bauthor{\bsnm{Machino}, \binits{H.}},
\bauthor{\bsnm{Yasutomi}, \binits{S.}},
\bauthor{\bsnm{Arakaki}, \binits{T.}},
\bauthor{\bsnm{Asada}, \binits{K.}},
\bauthor{\bsnm{Kaneko}, \binits{S.}}, \betal:
\batitle{Image segmentation of the ventricular septum in fetal cardiac
  ultrasound videos based on deep learning using time-series information}.
\bjtitle{Biomolecules}
\bvolume{10}(\bissue{11}),
\bfpage{1526}
(\byear{2020})
\end{barticle}
\endbibitem

%%% 57
\bibitem[\protect\citeauthoryear{Raissi et~al.}{2019}]{raissi2019physics}
\begin{barticle}
\bauthor{\bsnm{Raissi}, \binits{M.}},
\bauthor{\bsnm{Perdikaris}, \binits{P.}},
\bauthor{\bsnm{Karniadakis}, \binits{G.E.}}:
\batitle{Physics-informed neural networks: A deep learning framework for
  solving forward and inverse problems involving nonlinear partial differential
  equations}.
\bjtitle{Journal of Computational physics}
\bvolume{378},
\bfpage{686}--\blpage{707}
(\byear{2019})
\end{barticle}
\endbibitem

%%% 58
\bibitem[\protect\citeauthoryear{Karniadakis
  et~al.}{2021}]{karniadakis2021physics}
\begin{barticle}
\bauthor{\bsnm{Karniadakis}, \binits{G.E.}},
\bauthor{\bsnm{Kevrekidis}, \binits{I.G.}},
\bauthor{\bsnm{Lu}, \binits{L.}},
\bauthor{\bsnm{Perdikaris}, \binits{P.}},
\bauthor{\bsnm{Wang}, \binits{S.}},
\bauthor{\bsnm{Yang}, \binits{L.}}:
\batitle{Physics-informed machine learning}.
\bjtitle{Nature Reviews Physics}
\bvolume{3}(\bissue{6}),
\bfpage{422}--\blpage{440}
(\byear{2021})
\end{barticle}
\endbibitem

%%% 59
\bibitem[\protect\citeauthoryear{Wang et~al.}{2023}]{wang2023expert}
\begin{botherref}
\oauthor{\bsnm{Wang}, \binits{S.}},
\oauthor{\bsnm{Sankaran}, \binits{S.}},
\oauthor{\bsnm{Wang}, \binits{H.}},
\oauthor{\bsnm{Perdikaris}, \binits{P.}}:
An expert's guide to training physics-informed neural networks.
arXiv preprint arXiv:2308.08468
(2023)
\end{botherref}
\endbibitem

%%% 60
\bibitem[\protect\citeauthoryear{Hao et~al.}{2022}]{hao2022physics}
\begin{botherref}
\oauthor{\bsnm{Hao}, \binits{Z.}},
\oauthor{\bsnm{Liu}, \binits{S.}},
\oauthor{\bsnm{Zhang}, \binits{Y.}},
\oauthor{\bsnm{Ying}, \binits{C.}},
\oauthor{\bsnm{Feng}, \binits{Y.}},
\oauthor{\bsnm{Su}, \binits{H.}},
\oauthor{\bsnm{Zhu}, \binits{J.}}:
Physics-informed machine learning: A survey on problems, methods and
  applications.
arXiv preprint arXiv:2211.08064
(2022)
\end{botherref}
\endbibitem

%%% 61
\bibitem[\protect\citeauthoryear{Arzani et~al.}{2021}]{arzani2021uncovering}
\begin{botherref}
\oauthor{\bsnm{Arzani}, \binits{A.}},
\oauthor{\bsnm{Wang}, \binits{J.-X.}},
\oauthor{\bsnm{D'Souza}, \binits{R.M.}}:
Uncovering near-wall blood flow from sparse data with physics-informed neural
  networks.
Physics of Fluids
\textbf{33}(7)
(2021)
\end{botherref}
\endbibitem

%%% 62
\bibitem[\protect\citeauthoryear{Farea et~al.}{2024}]{farea2024understanding}
\begin{barticle}
\bauthor{\bsnm{Farea}, \binits{A.}},
\bauthor{\bsnm{Yli-Harja}, \binits{O.}},
\bauthor{\bsnm{Emmert-Streib}, \binits{F.}}:
\batitle{Understanding physics-informed neural networks: Techniques,
  applications, trends, and challenges}.
\bjtitle{AI}
\bvolume{5}(\bissue{3}),
\bfpage{1534}--\blpage{1557}
(\byear{2024})
\end{barticle}
\endbibitem

%%% 63
\bibitem[\protect\citeauthoryear{Huang and Wang}{2022}]{huang2022applications}
\begin{barticle}
\bauthor{\bsnm{Huang}, \binits{B.}},
\bauthor{\bsnm{Wang}, \binits{J.}}:
\batitle{Applications of physics-informed neural networks in power systems-a
  review}.
\bjtitle{IEEE Transactions on Power Systems}
\bvolume{38}(\bissue{1}),
\bfpage{572}--\blpage{588}
(\byear{2022})
\end{barticle}
\endbibitem

%%% 64
\bibitem[\protect\citeauthoryear{Sharma et~al.}{2023}]{sharma2023review}
\begin{barticle}
\bauthor{\bsnm{Sharma}, \binits{P.}},
\bauthor{\bsnm{Chung}, \binits{W.T.}},
\bauthor{\bsnm{Akoush}, \binits{B.}},
\bauthor{\bsnm{Ihme}, \binits{M.}}:
\batitle{A review of physics-informed machine learning in fluid mechanics}.
\bjtitle{Energies}
\bvolume{16}(\bissue{5}),
\bfpage{2343}
(\byear{2023})
\end{barticle}
\endbibitem

%%% 65
\bibitem[\protect\citeauthoryear{Conti et~al.}{2024}]{conti2024multi}
\begin{barticle}
\bauthor{\bsnm{Conti}, \binits{P.}},
\bauthor{\bsnm{Guo}, \binits{M.}},
\bauthor{\bsnm{Manzoni}, \binits{A.}},
\bauthor{\bsnm{Frangi}, \binits{A.}},
\bauthor{\bsnm{Brunton}, \binits{S.L.}},
\bauthor{\bsnm{Nathan~Kutz}, \binits{J.}}:
\batitle{Multi-fidelity reduced-order surrogate modelling}.
\bjtitle{Proceedings of the Royal Society A}
\bvolume{480}(\bissue{2283}),
\bfpage{20230655}
(\byear{2024})
\end{barticle}
\endbibitem

%%% 66
\bibitem[\protect\citeauthoryear{Taebi}{2022}]{taebi2022deep}
\begin{barticle}
\bauthor{\bsnm{Taebi}, \binits{A.}}:
\batitle{Deep learning for computational hemodynamics: A brief review of recent
  advances}.
\bjtitle{Fluids}
\bvolume{7}(\bissue{6}),
\bfpage{197}
(\byear{2022})
\end{barticle}
\endbibitem

%%% 67
\bibitem[\protect\citeauthoryear{Olakorede}{2022}]{olakorede2022physics}
\begin{botherref}
\oauthor{\bsnm{Olakorede}, \binits{I.}}:
Physics-based artificial intelligence for atrial electrophysiological model
(2022)
\end{botherref}
\endbibitem

%%% 68
\bibitem[\protect\citeauthoryear{Sarabian et~al.}{2022}]{sarabian2022physics}
\begin{barticle}
\bauthor{\bsnm{Sarabian}, \binits{M.}},
\bauthor{\bsnm{Babaee}, \binits{H.}},
\bauthor{\bsnm{Laksari}, \binits{K.}}:
\batitle{Physics-informed neural networks for brain hemodynamic predictions
  using medical imaging}.
\bjtitle{IEEE transactions on medical imaging}
\bvolume{41}(\bissue{9}),
\bfpage{2285}--\blpage{2303}
(\byear{2022})
\end{barticle}
\endbibitem

%%% 69
\bibitem[\protect\citeauthoryear{Panneerselvam
  et~al.}{2026}]{panneerselvam2026toward}
\begin{barticle}
\bauthor{\bsnm{Panneerselvam}, \binits{N.K.}},
\bauthor{\bsnm{Mummaneni}, \binits{G.}},
\bauthor{\bsnm{Roncali}, \binits{E.}}:
\batitle{Toward digital twins for optimal radioembolization}.
\bjtitle{PET clinics}
\bvolume{21}(\bissue{1}),
\bfpage{153}--\blpage{167}
(\byear{2026})
\end{barticle}
\endbibitem

%%% 70
\bibitem[\protect\citeauthoryear{Lydon et~al.}{2025}]{lydon2025physics}
\begin{botherref}
\oauthor{\bsnm{Lydon}, \binits{H.}},
\oauthor{\bsnm{Kazemi}, \binits{M.}},
\oauthor{\bsnm{Bishop}, \binits{M.}},
\oauthor{\bsnm{Paoletti}, \binits{N.}}:
Physics-informed neural operators for cardiac electrophysiology.
arXiv preprint arXiv:2511.08418
(2025)
\end{botherref}
\endbibitem

%%% 71
\bibitem[\protect\citeauthoryear{Herrero~Martin et~al.}{2022}]{herrero2022ep}
\begin{barticle}
\bauthor{\bsnm{Herrero~Martin}, \binits{C.}},
\bauthor{\bsnm{Oved}, \binits{A.}},
\bauthor{\bsnm{Chowdhury}, \binits{R.A.}},
\bauthor{\bsnm{Ullmann}, \binits{E.}},
\bauthor{\bsnm{Peters}, \binits{N.S.}},
\bauthor{\bsnm{Bharath}, \binits{A.A.}},
\bauthor{\bsnm{Varela}, \binits{M.}}:
\batitle{Ep-pinns: Cardiac electrophysiology characterisation using
  physics-informed neural networks}.
\bjtitle{Frontiers in Cardiovascular Medicine}
\bvolume{8},
\bfpage{768419}
(\byear{2022})
\end{barticle}
\endbibitem

%%% 72
\bibitem[\protect\citeauthoryear{Meng et~al.}{2025}]{meng2025physics}
\begin{barticle}
\bauthor{\bsnm{Meng}, \binits{C.}},
\bauthor{\bsnm{Griesemer}, \binits{S.}},
\bauthor{\bsnm{Cao}, \binits{D.}},
\bauthor{\bsnm{Seo}, \binits{S.}},
\bauthor{\bsnm{Liu}, \binits{Y.}}:
\batitle{When physics meets machine learning: A survey of physics-informed
  machine learning}.
\bjtitle{Machine Learning for Computational Science and Engineering}
\bvolume{1}(\bissue{1}),
\bfpage{20}
(\byear{2025})
\end{barticle}
\endbibitem

%%% 73
\bibitem[\protect\citeauthoryear{Barbiero et~al.}{2021}]{barbiero2021graph}
\begin{barticle}
\bauthor{\bsnm{Barbiero}, \binits{P.}},
\bauthor{\bsnm{Vinas~Torne}, \binits{R.}},
\bauthor{\bsnm{Li{\'o}}, \binits{P.}}:
\batitle{Graph representation forecasting of patient's medical conditions:
  toward a digital twin}.
\bjtitle{Frontiers in genetics}
\bvolume{12},
\bfpage{652907}
(\byear{2021})
\end{barticle}
\endbibitem

%%% 74
\bibitem[\protect\citeauthoryear{Paul et~al.}{2024}]{paul2024systematic}
\begin{barticle}
\bauthor{\bsnm{Paul}, \binits{S.G.}},
\bauthor{\bsnm{Saha}, \binits{A.}},
\bauthor{\bsnm{Hasan}, \binits{M.Z.}},
\bauthor{\bsnm{Noori}, \binits{S.R.H.}},
\bauthor{\bsnm{Moustafa}, \binits{A.}}:
\batitle{A systematic review of graph neural network in healthcare-based
  applications: Recent advances, trends, and future directions}.
\bjtitle{IEEE Access}
\bvolume{12},
\bfpage{15145}--\blpage{15170}
(\byear{2024})
\end{barticle}
\endbibitem

%%% 75
\bibitem[\protect\citeauthoryear{Shukla et~al.}{2022}]{shukla2022scalable}
\begin{barticle}
\bauthor{\bsnm{Shukla}, \binits{K.}},
\bauthor{\bsnm{Xu}, \binits{M.}},
\bauthor{\bsnm{Trask}, \binits{N.}},
\bauthor{\bsnm{Karniadakis}, \binits{G.E.}}:
\batitle{Scalable algorithms for physics-informed neural and graph networks}.
\bjtitle{Data-Centric Engineering}
\bvolume{3},
\bfpage{24}
(\byear{2022})
\end{barticle}
\endbibitem

%%% 76
\bibitem[\protect\citeauthoryear{Iacovelli et~al.}{2024}]{iacovelli2024novel}
\begin{bchapter}
\bauthor{\bsnm{Iacovelli}, \binits{A.}},
\bauthor{\bsnm{Pegolotti}, \binits{L.}},
\bauthor{\bsnm{Salvador}, \binits{M.}},
\bauthor{\bsnm{Stoppa}, \binits{E.}},
\bauthor{\bsnm{Santambrogio}, \binits{M.D.}},
\bauthor{\bsnm{Marsden}, \binits{A.}}:
\bctitle{A novel lstm and graph neural networks approach for cardiovascular
  simulations}.
In: \bbtitle{2024 IEEE EMBS International Conference on Biomedical and Health
  Informatics (BHI)},
pp. \bfpage{1}--\blpage{8}
(\byear{2024}).
\bcomment{IEEE}
\end{bchapter}
\endbibitem

%%% 77
\bibitem[\protect\citeauthoryear{Oloulade et~al.}{2021}]{oloulade2021graph}
\begin{barticle}
\bauthor{\bsnm{Oloulade}, \binits{B.M.}},
\bauthor{\bsnm{Gao}, \binits{J.}},
\bauthor{\bsnm{Chen}, \binits{J.}},
\bauthor{\bsnm{Lyu}, \binits{T.}},
\bauthor{\bsnm{Al-Sabri}, \binits{R.}}:
\batitle{Graph neural architecture search: A survey}.
\bjtitle{Tsinghua Science and Technology}
\bvolume{27}(\bissue{4}),
\bfpage{692}--\blpage{708}
(\byear{2021})
\end{barticle}
\endbibitem

%%% 78
\bibitem[\protect\citeauthoryear{You et~al.}{2020}]{you2020design}
\begin{barticle}
\bauthor{\bsnm{You}, \binits{J.}},
\bauthor{\bsnm{Ying}, \binits{Z.}},
\bauthor{\bsnm{Leskovec}, \binits{J.}}:
\batitle{Design space for graph neural networks}.
\bjtitle{Advances in Neural Information Processing Systems}
\bvolume{33},
\bfpage{17009}--\blpage{17021}
(\byear{2020})
\end{barticle}
\endbibitem

%%% 79
\bibitem[\protect\citeauthoryear{Kuang et~al.}{2024}]{kuang2024med}
\begin{barticle}
\bauthor{\bsnm{Kuang}, \binits{K.}},
\bauthor{\bsnm{Dean}, \binits{F.}},
\bauthor{\bsnm{B~Jedlicki}, \binits{J.}},
\bauthor{\bsnm{Ouyang}, \binits{D.}},
\bauthor{\bsnm{Philippakis}, \binits{A.}},
\bauthor{\bsnm{Sontag}, \binits{D.}},
\bauthor{\bsnm{Alaa}, \binits{A.M.}}:
\batitle{Med-real2sim: Non-invasive medical digital twins using
  physics-informed self-supervised learning}.
\bjtitle{Advances in Neural Information Processing Systems}
\bvolume{37},
\bfpage{5757}--\blpage{5788}
(\byear{2024})
\end{barticle}
\endbibitem

%%% 80
\bibitem[\protect\citeauthoryear{Rieke et~al.}{2020}]{rieke2020future}
\begin{barticle}
\bauthor{\bsnm{Rieke}, \binits{N.}},
\bauthor{\bsnm{Hancox}, \binits{J.}},
\bauthor{\bsnm{Li}, \binits{W.}},
\bauthor{\bsnm{Milletari}, \binits{F.}},
\bauthor{\bsnm{Roth}, \binits{H.R.}},
\bauthor{\bsnm{Albarqouni}, \binits{S.}},
\bauthor{\bsnm{Bakas}, \binits{S.}},
\bauthor{\bsnm{Galtier}, \binits{M.N.}},
\bauthor{\bsnm{Landman}, \binits{B.A.}},
\bauthor{\bsnm{Maier-Hein}, \binits{K.}}, \betal:
\batitle{The future of digital health with federated learning}.
\bjtitle{NPJ digital medicine}
\bvolume{3}(\bissue{1}),
\bfpage{119}
(\byear{2020})
\end{barticle}
\endbibitem

\end{thebibliography}


\end{document}